\documentclass[twocolumn,amsmath,amssymb,aps,prd,floatfix,10pt,showpacs]{revtex4-1}

\usepackage{graphicx}
\usepackage{dcolumn}
\usepackage{amsmath}
\usepackage{amsfonts}
\usepackage{hyperref}

\hypersetup{%
pdftitle = {Are the Dressed Gluon and Ghost Propagators in the Landau Gauge presently determined in the confinement regime of QCD?},
pdfsubject = {QCD},
pdfkeywords = {QCD, SDE, DSE, Gluon, Ghost, Dressing, Non-perturbative},
pdfauthor = {Michael R. Pennington, David J. Wilson},
colorlinks = {true},
filecolor = {black},
linkcolor = {black},
menucolor = {black},
citecolor = {black},
urlcolor = {black},
}{}

\newcommand{\Gl}{\ensuremath{\mathcal{G}\ell}}
\newcommand{\Gh}{\ensuremath{\mathcal{G}h}}

\begin{document}
\title{Are the Dressed Gluon and Ghost Propagators in the Landau Gauge\\ presently determined in the confinement regime of QCD?}

\author{M.R. Pennington$^{a}$} 
\author{D.J. Wilson$^{b}$}

\affiliation{
$^a$ Theory Center, Thomas Jefferson National Accelerator Facility, 12000 Jefferson Avenue, Newport News, Virginia 23606, USA.\\
$^b$ Argonne National Laboratory, Argonne, Illinois 60439, USA.}
\date{\today}

\begin{abstract}
The Gluon and Ghost propagators in Landau gauge QCD are investigated using the Schwinger-Dyson equation approach. Working in Euclidean spacetime, we solve for these propagators using a selection of vertex inputs, initially for the ghost equation alone and then for both propagators simultaneously. The results are shown to be highly sensitive to the choices of vertices. We favor the infrared finite ghost solution from studying the ghost equation alone where we argue for a specific unique solution. In order to solve this simultaneously with the gluon using a dressed-one-loop truncation, we find that a non-trivial full ghost-gluon vertex is required in the vanishing gluon momentum limit. The self-consistent solutions we obtain correspond to having a mass-like term in the gluon propagator dressing, in agreement with similar studies supporting the long held proposal of Cornwall.
\end{abstract}


\maketitle

\section{Introduction}

The Gluon Schwinger-Dyson equation and its gauge fixing counterpart, the Ghost Schwinger-Dyson equation, are principal tools in investigating the theoretical properties of strongly coupled QCD. The transition from the weak coupling regime at large momenta, where perturbation theory applies, down to strong coupling at small momenta gives rise to the key non-perturbative phenomena of confinement and dynamical chiral symmetry breaking. The latter endows quarks, and perhaps gluons, with mass. These are not physical pole masses, but rather effective Euclidean masses dynamically generated by non-perturbative interactions~\cite{Cornwall:1981zr,Mandula:1987rh}. The main reason for studying these propagators is that they are a necessary input for the Bethe-Salpeter (BSE) and Faddeev equations used for calculating physical quantities, such as hadron masses and form-factors~\cite{Roberts:1994dr,Chang:2011vu}.

The infinite tower of Schwinger-Dyson equations (SDEs), and its necessary truncations in QCD, can be a troublesome beast. This paper is intended to stimulate discussion about the assumptions commonly made and the solutions to which these lead. We believe that important issues remain unaddressed and these are outlined below. 

The solutions provided by the SDEs may be compared with the results of complementary techniques, such as Lattice QCD. A benefit of this is that the theoretical strengths and weaknesses of each method fall in different areas. Both  may be formulated in precisely the same way, using the same gauge and freely varying the quark content. While lattice calculations are inevitably restricted to larger quark mass, an advantage of the SDE method is that there are no major computational obstacles to extending these to the physical light quarks with masses of $\cal{O}$(MeV).

The paper is organised as follows, in Sec. 2, we introduce the Schwinger-Dyson equations for the gluon and ghost system and proceed to give details regarding truncations. In Sec. 3 we solve the ghost equation for simple truncations and show that solutions may be robustly obtained for a wide range of vertex inputs. In Sec. 4 we outline the gluon equation and some of the issues we deem important that deserve immediate attention. We present simultaneous solutions for the gluon and ghost propagators for a range of assumptions to indicate the level of precision currently available. In Sec. 5 we compare our results to those obtained using Lattice QCD simulations. Section~6 briefly gives our conclusions and outlook.
 
\section{Schwinger-Dyson equations for Gluons and Ghosts}

\begin{figure}[!t]
  \begin{center}
  \includegraphics[width=0.4\textwidth]{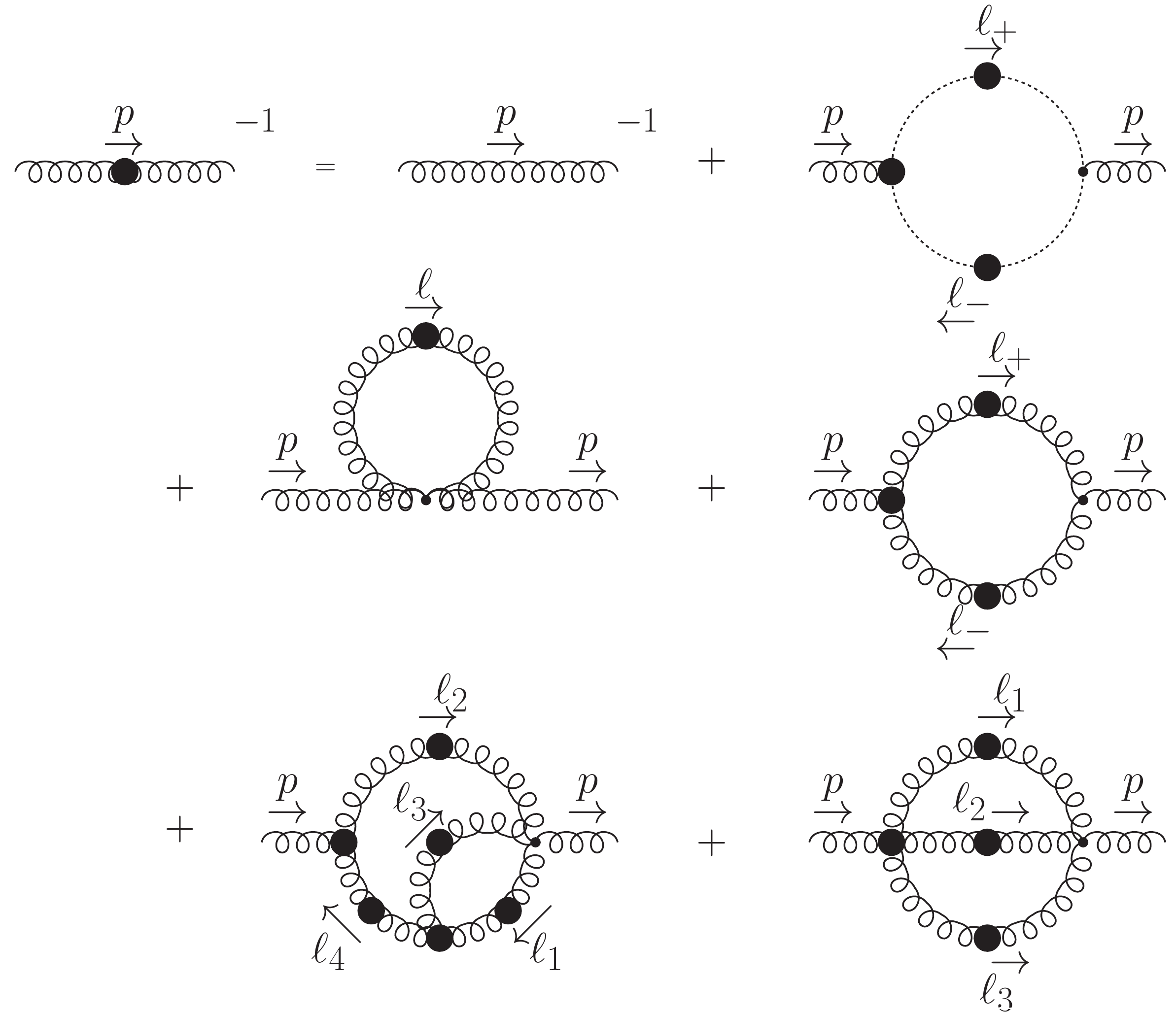}   
  \end{center}
  \caption{The Schwinger--Dyson Equation for the Gluon in the absence of quarks.}
  \label{fig_sdefullgluon}
\end{figure}

The full equation for the gluon propagator, in the absence of quarks, is represented in Fig.~\ref{fig_sdefullgluon}. Without approximations, this is given by,
%
%
\begin{align}
D_{\mu\nu}^{ab}(p)^{-1}&=\,D^{ab,(0)}_{\mu\nu}(p)^{-1} + \Pi^{ab}_{2c\,\mu\nu}(p)+\Pi^{ab}_{1g\,\mu\nu}(p)\nonumber\\
&+\Pi^{ab}_{2g\,\mu\nu}(p)+\Pi^{ab}_{3g\,\mu\nu}(p)+\Pi^{ab}_{4g\,\mu\nu}(p),
\label{eq_gluonsde_full}
\end{align}
where latin indices denote colour and greek letters denote the Lorentz suffixes. The loop integrations $\Pi(p)$ are labeled by the number of particles in the loop with $c$ corresponding to ghosts and $g$ corresponding to gluons. Eq.~(\ref{eq_gluonsde_full}) is represented graphically in Fig.~\ref{fig_sdefullgluon}. The momenta flowing through the internal propagators are arbitrary. However we use a symmetric routing for the one-loop diagrams with $\ell_{\pm}=\ell\pm p/2$, where $\ell$ is the loop integration momentum and $p$ is the external momentum that is divided equally through each propagator. When working with a large finite cutoff as is often done numerically, then a symmetric arrangement is often essential to preserve translation invariance.

\begin{figure}[!b]
  \begin{center}
  \includegraphics[width=0.4\textwidth]{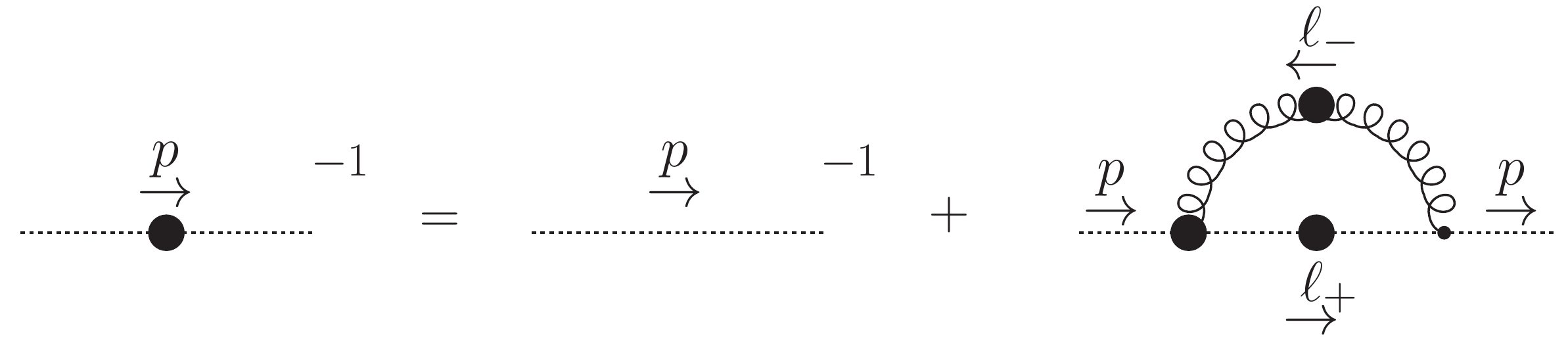}   
  \end{center}
  \caption{The Schwinger--Dyson Equation for the Ghost.}
  \label{fig_sdefullghost}
\end{figure}

We work in covariant gauges. Then neglecting quarks, the gluon equation only couples to ghosts, and these have their own Schwinger-Dyson equation, which is depicted in Fig.~\ref{fig_sdefullghost} and given by,
\begin{align}
D(p)^{-1}=D^{(0)}(p)^{-1}+\Pi_{gc}(p),
\end{align}
where $\Pi_{gc}$ contains just one simple loop integration.

Solving these equations depends non-trivially on the vertices, which in turn depend on higher $n$-point Green's functions and therein lies the key issue. Truncation of this sequence of dependent Green's functions is necessary to be able to make predictions. However truncating has the potential to introduce errors and violate the physical properties of the theory.

\subsection{Sketching the general features of the equations}
The interplay of the various terms in the equations and identifying which terms are important will be our primary concern. For the most part we will adopt a truncation in which only one loop dressing of the gluon propagator will be considered. This is a favorite truncation scheme~\cite{Atkinson:1997tu,vonSmekal:1997vx,Lerche:2002ep,Alkofer:2003jr,Fischer:2003zc,Aguilar:2008xm,Fischer:2008uz}. There are then four quantities of interest seen in Fig.~1: the two propagator dressings that the equations should determine, and the two unknown vertices that are required inputs. There are important interrelations between these terms constrained by the structure of the SDEs and the Ward-Slavnov-Taylor identities (WSTIs).

The propagators are remarkably simple. The ghost and gluon dressings are just functions of $p^{2}$ with the gluon dressing multiplied by a tensor structure transverse to its momentum $p$. In the Landau gauge these are,
\begin{align}
\mathcal{D}^{ab}_{\mu\nu}(p^{2})&=\delta^{ab}\frac{\Gl(p^{2})}{p^{2}}\left(g_{\mu\nu}-\frac{p_{\mu}p_{\nu}}{p^{2}}\right)\\
\mathcal{D}^{ab}(p^{2})&=-\delta^{ab}\frac{\Gh(p^{2})}{p^{2}}
\end{align}
where $\mathcal{D}_{\mu\nu}$ is the full gluon propagator, $\mathcal{D}$ is the ghost propagator. The $a,b$ indices relate to the color carried by each propagator. The functions $\Gl$ and $\Gh$ are the respective dressing functions, for which we will be solving later. These functions contain all of the non-perturbative physics of these two Green's functions.

The two propagator dressing functions are determined from the various quantities that make up their equations, which are represented diagrammatically in Fig.~\ref{fig_sdefullgluon} and Fig.~\ref{fig_sdefullghost}. The triple-gluon vertex only appears in the gluon equation. However the two propagators and the ghost-gluon vertex appear in both equations. While the dressing functions are not expected to change sign over the whole Euclidean momentum region, the different contributions from the loop diagrams can have different signs, and indeed they do. The ghost-loop has an additional $(-1)$ owing to the antisymmetry of the ghost field. 

The gluon propagator dressing determined in the momentum subtraction scheme is given by,
\begin{align}
\Gl(p^{2})^{-1}=\Gl(\mu^{2})^{-1} + \Pi_{2c}(p^{2},\mu^{2}) + \Pi_{2g}(p^{2},\mu^{2})
\label{eqglsub}
\end{align}
where $\mu^{2}$ is the subtractive-renormalization point, $\Pi_{2c}(p^{2},\mu^{2})$ is the ghost-loop contribution subtracted from itself at $\mu^{2}$, similarly for $\Pi_{2g}(p^{2},\mu^{2})$ which is the gluon loop contribution. The precise details may be found in the appendices.

In the definitions we use, the gluon propagator dressing function $\Gl(p^{2})$ is positive everywhere. However, the dressed-gluon-loop diagram gives a negative contribution in the momentum subtraction scheme, below the subtraction point. This is important because it tells us that the two other terms on the right of Eq.~(\ref{eqglsub}), $\Gl(\mu^{2})$ and $\Pi_{2c}(p^{2},\mu^{2})$ must add to give an overall positive contribution of greater magnitude than that of $\Pi_{2g}(p^{2},\mu^{2})$. Now in the perturbative region $\Gl(\mu^{2})=1$ is more than large enough to ensure that $\Gl(p^{2})$ remains positive, but as we evolve down to infrared (IR) momenta, then the two one-loop contributions are both large. Quite generically for many sensible vertices, the ghost-loop term must then give a large positive contribution in order to be able to solve the equation. A dichotomy of explanations exist: the large contribution either arises from a singular ghost propagator, or a non-trivial 
 ghost-gluon vertex. Despite this distinction, little change is observed in the gluon dressing function since the overall contribution from the ghost loop remains similar.

The full ghost-gluon vertex has two structures that may be written as follows,
\begin{align}
\Gamma_{\mu}(k,p,q)=igf^{abc}\left[\,q_{\mu}\,\alpha(k,p,q)\,+\,k_{\mu}\,\beta(k,p,q)\,\right]\; ,
\label{eq_gccgen}
\end{align}
where the momenta are defined in Fig.~\ref{fig_gccdiag} and $\alpha$ is the non-perturbative function dressing the outgoing ghost momentum and $\beta$ is the non-perturbative function dressing the outgoing gluon term. This can be arranged in a number of ways since the momenta are related by $k=p-q$. In the ghost equation the term $\beta$ does not contribute in Landau gauge, however in a properly projected gluon the $\beta$ term can be important. It appears in the ghost-loop, as we emphasise later.

The ghost-gluon scattering kernel $\tilde{\Gamma}_{\mu\nu}$ is an important quantity since it appears in dressing both vertices. It is directly related to the ghost-gluon vertex,
\begin{equation}
\Gamma_{\mu}(k,p,q)=igf^{abc}\,q_{\nu}\tilde{\Gamma}_{\nu\mu}(k,p,q)\; .
\end{equation}
Hence the bare version is $\tilde{\Gamma}_{\mu\nu}=g_{\mu\nu}$ in order to yield the tree-level vertex. The same function also appears in the WSTI for the triple-gluon vertex, which is a feature of the interconnections between terms in a gauge theory. The precise relation will probably be necessary to remove all gauge-dependence from the results. We do not give further details here as these are fully covered in~\cite{vonSmekal:1997vx,Aguilar:2011ux} and references therein.

\subsection{Gauge fixing}

We work in Landau gauge since it happens to be the most theoretically appealing for several reasons. An obvious and oft-suggested alternative is to use a gauge defined by an axial vector $n_\mu$, where ghosts do not appear. However these have other drawbacks, not the least of which is the breaking of Lorentz symmetry, and in particular in numerical studies where $1/(n.p)$ terms are a complication. Additionally, direct comparison to Lattice QCD is less straightforward in these gauges. The radiative corrections to the Gluon propagator are always transverse in the external momentum due to the Slavnov-Taylor identity, and in the Landau gauge the Gluon itself is also transverse, which is a simplifying feature.
Furthermore, there are theoretical statements regarding confinement, gauge fixing and the vanishing momentum limit of QCD. These are usually formulated in Landau gauge and some details are given below.

Ultimately we hope that these methods will be useful in making physical predictions where the gauge dependence must drop out as in perturbation theory. However, the gluon propagator dressing function is a gauge dependent object and one expects it will combine with all of the other gauge dependent functions in making gauge-independent physical observables.

A deeper analysis of the gauge fixing procedure is required on theoretical grounds, since the Faddeev-Popov procedure may not be sufficient non-perturbatively. Earlier studies by Gribov, Kugo, Ojima, and Zwanziger have received refinement in recent years in light of the serious debate over the behavior of the propagators in the vanishing momentum limit~\cite{Gribov:1977wm,Zwanziger:1991gz,Zwanziger:1992qr,Zwanziger:1993dh}. Formulated in Landau gauge, the Gribov-Zwanziger gauge fixing procedure considers the technical issues related to gauge field Gribov copies and restriction of the gauge field space to select one representative configuration. The broad conclusion at present is that the dressed gluon propagator carrying momentum $p$ is suppressed as $p$ vanishes, with respect to its bare counterpart. Moreover, many studies find a propagator that goes to a constant, or equivalently a propagator dressing function that is proportional to $p^{2}$ in the small $p^{2}$ limit.

Coupled with the earlier conclusions of Kugo and Ojima~\cite{Kugo:1979gm} regarding a singular ghost dressing being a signal of confinement, SDE practioners were led to search for a particular class of solution where the leading powers in the IR of the ghost and gluon propagators are directly linked. More details of this class of solution will be outlined below, although the key feature is the singular ghost dressing function. Recently, new studies combining these two earlier studies, and also additional refinements using Stochastic quantization techniques, have found that a finite ghost dressing solution is certainly allowed and may even be preferred~\cite{Zwanziger:2002ia,Dudal:2008sp,Kondo:2009gc,Kondo:2009wk,Dudal:2011gd}.

\subsection{Existing Truncation Schemes and Solutions}
We now turn to the act of truncation and briefly summarise some of the ideas previously considered for these equations.
Most truncation schemes for solving the gluon equation are motivated by the practicalities of finding solutions. For example, in all but one of the references in the literature~\cite{Bloch:2003yu} the dressed-two-loop diagrams of Fig.~\ref{fig_sdefullgluon} are entirely neglected. This is not due to some convenient power counting scheme, but rather because their inclusion is both numerically challenging and computationally intensive. Clearly in the perturbative ultra-violet (UV) regime,  these terms are subleading. However in the intermediate and vanishing momentum limit, their contribution is essentially unknown. Na\"{i}vely, one might expect that if the gluon is heavily suppressed in the small momentum limit then these terms may be unimportant. In this case of an IR suppressed gluon, the most important region for these terms would be the mid-momentum region (\emph{i.e.} $p \sim 1$ GeV) as the leading perturbative logarithm  reaches a maximum and turns 
over in the non-perturbative region. For a gluon propagator that is constant in the IR (or equivalently a gluon propagator dressing function that goes as $p^{2}$), then the situation is less clear. The answer depends on the vertex ans\"atze adopted and the full solution of the gluon dressing function. The mid-momentum region is also the most physically relevant so eventually these terms must be properly included.

A recent one-loop perturbative study, in which an effective gluon mass term is artificially introduced, has found good agreement~\cite{Tissier:2011ey} with results from Lattice QCD. This differs from a typical SDE study since the mass is inserted by hand and iterations are not performed. Otherwise the equations are very closely related. In order to perform iterations required for self-consistent solutions (see Appendix A), some term in the gluon equation is required to dynamically generate the mass term. In the SDEs, this type of contribution could come from any of the loops in the gluon equation including the two loop graphs.

Nevertheless, we proceed by neglecting these two-loop dressing terms and turn to the remaining two unknown vertices that appear in the ghost and gluon equations. The ideal situation would be that the propagators depend only weakly on the ans\"atze for these vertices, and it is the propagator itself and the structure of the SDE that determines the result. In studies in QED the situation is indeed close to this ideal, since the Ward-Green-Takahashi identity determines the key vertex dressings in terms of inverse propagators~\cite{Ball:1980ay,Ball:1980ax,Curtis:1990zs,Bashir:1997qt,Bashir:2007qq,Kizilersu:2009kg}. QCD is inevitably more complicated. The Ward-Slavnov-Taylor Identities that relate vertices and propagators do not admit such simple solutions as in QED, and for the full vertex there is a complex interplay between the gluon and ghost sectors.

\begin{figure}[tbh]
  \begin{center}
  \includegraphics[width=0.2\textwidth]{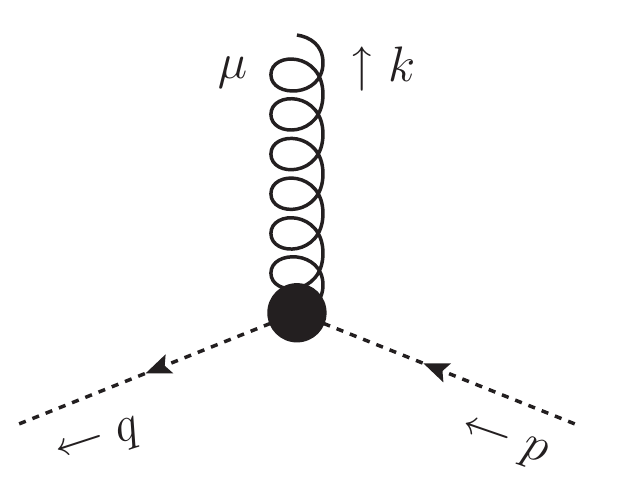}   
  \end{center}
  \caption{The ghost-gluon vertex indicating the momentum definition we adopt, the outgoing ghost momentum is $q$ and the gluon momentum is $k$.}
  \label{fig_gccdiag}
\end{figure}

In QCD the starting point has often been bare vertices, or vertices dressed by simple ratios of dressing functions~\cite{BarGadda:1979cz,Brown:1988bm,Brown:1988bn,Roberts:1994dr,Atkinson:1997tu,Alkofer:2003jr,Alkofer:2004it}. In particular, one option is to use the bare ghost-gluon vertex in place of the full vertex,
\begin{equation}
   \Gamma^{abc}_{\mu}(k,p,q)\to\Gamma^{(0)\,abc}_{\mu}(k,p,q)=igf^{abc}q_{\mu}\,,
   \label{eq_baregcc}
\end{equation}
which at first sight seems like an over-simplification, since this cannot be correct even at the perturbative one-loop order~\cite{Celmaster:1979km,Davydychev:1996pb}, let alone non-perturbatively. However, it is expected that the ghost-gluon vertex should be relatively simple in the Landau gauge, by appealing to Taylor's non-renormalization theorem \cite{Taylor:1971ff} which is valid in carefully chosen renormalization schemes \cite{Boucaud:2005ce,Boucaud:2008ky}. The full vertex is also expected to reduce to its bare form for vanishing incoming ghost momenta~\cite{Taylor:1971ff,Lerche:2002ep,Boucaud:2005ce},
\begin{equation}
   \lim_{p\to0}\Gamma^{abc}_{\mu}(k,p,q)=\Gamma^{(0)\,abc}_{\mu}(k,p,q),
\end{equation}
the momentum definition used throughout being indicated in Fig.~\ref{fig_gccdiag}. This simple vertex appears to work for the infinite IR ghost dressing function~\cite{Alkofer:2003jr,Bloch:2003yu}. However in the finite solutions a problem arises. In practical terms, the large contribution from the infinite ghost must be replaced by a large contribution from the vertices in order for the gluon dressing function to remain positive, and for self-consistent solutions to be obtained. There are two forms given in the literature that accomplish this. Firstly, a vertex that is transverse to the gluon momentum in the IR is suggested as a possible solution~\cite{Lerche:2002ep,Fischer:2008uz}. This is motivated by solutions in which the ghost is dominant in the IR region. Then the ghost loop controls corrections to the gluon propagator, see Fig.~1, which must then be transverse on its own in this regime. This can be achieved by making the ghost-gluon vertex 
transverse to the gluon momentum in the IR. A form that achieves this is~\cite{Lerche:2002ep,Fischer:2008uz},
\begin{equation}
\Gamma_{\mu}^{(1)}=igf^{abc}\left(q_{\mu} - k_{\mu}\frac{k.q}{k^{2}}\mathcal{F}_\mathrm{IR}(k,p,q)\right).
\label{eq_gcctransv}
\end{equation}
where $\mathcal{F}_\mathrm{IR}$ is a smoothed step function defined in Eq.~(\ref{eq_fir}) that switches on this behavior in the IR, but ensures it does not affect the UV. $\mathcal{F}_{\mathrm{IR}}=1$ in the IR and must vanish in the UV in order to reproduce the perturbative results. To achieve this, the function $\mathcal{F}_\mathrm{IR}$ contains a free parameter that controls the momentum of the switchover point. This has to be fixed and some sensitivity to its value will be present in the solutions. This extra term has no effect in the ghost equation, where the transverse gluon is contracted with this vertex, and so it automatically drops out. It can however appear in the gluon equation as we demonstrate below. The superscript label $(1)$ in Eq.~(10) will be used to refer to this vertex later.

The second form considered in the literature involves inserting massless poles into the vertex that give an IR enhancement~\cite{Schwinger:1962tn,Schwinger:1962tp,Jackiw:1973tr}. This is in an alternative truncation where the Feynman diagrams are collected into individually transverse groups using pinch-technique and background field method rearrangements~\cite{Aguilar:2008xm,Binosi:2009qm}. There  is not a direct comparison of such a vertex insertion in this formulation with the standard SDEs we use. We can however make similar considerations and investigate the effects of massless poles in the vertex.

Several different choices are made for the triple-gluon vertex in the literature, which is perhaps a little suprising since sensible WSTI solutions are available. However these depend on an unknown contribution from the ghost-gluon vertex. When solved self-consistently, the SDE solutions should be matched on to the resummed leading-logarithm from perturbation theory. However using symmetric triple-gluon vertices, it has been found that the anomalous dimension of this logarithm is not exactly reproduced~\cite{Atkinson:1997tu}.

Proposed solutions to this issue include solving an equation relating the vertex dressing and the anomalous dimension in the perturbative region in order to choose a vertex dressing that gives the correct outcome. Typically these vertices sacrifice Bose-symmetry between the internally contracted and external legs of the vertex. When a dressed \emph{symmetric} vertex is considered, this diagram then contributes in the IR region of the gluon equation. This in turn requires an interplay between the ghost and gluon loops in the IR for the gluon equation and arguments about an individually transverse ghost-diagram may not necessarily be valid \cite{Lerche:2002ep,Fischer:2008uz}.

\begin{figure}[thb]
  \begin{center}
  \includegraphics[width=0.45\textwidth]{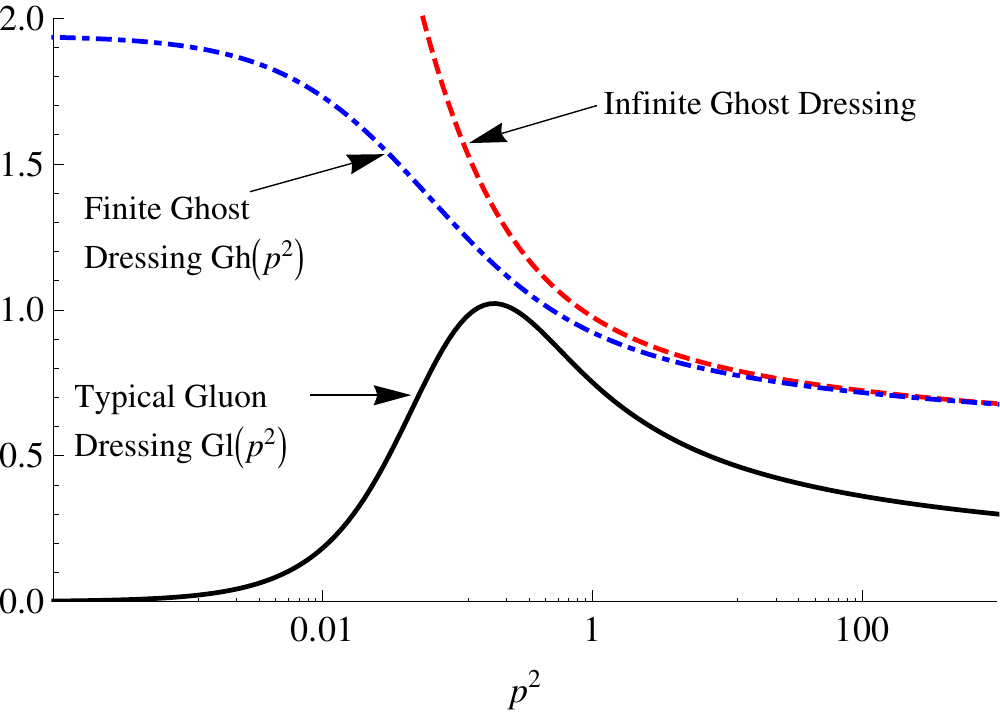}   
  \end{center}
  \caption{A sketch of typical dressing functions. The momentum scale is arbitrary, but can be thought of as $\sim 1$ GeV.}
  \label{fig_initialplot}
\end{figure}%

There are two broad classes of solution in existence and these have been strongly debated \cite{Aguilar:2008xm,Fischer:2008uz,Cucchieri:2011um}. The gluon, shown in Fig.~4, is qualitatively similar in most recent studies. However the ghost in the IR can vary wildly: from a finite value to infinity. We consider the ghost equation in detail in the next section. There are also solutions from Lattice QCD that are produced in the absence of quarks. At present these point to a finite ghost propagator dressing function~\cite{Bogolubsky:2009dc}. We give an example of some solutions in Fig.~\ref{fig_initialplot} including a typical gluon dressing with the logarithm from perturbation theory at large $p^{2}$ and a suppression at small $p^{2}$. The ghost is less certain and a wide range of solutions may be found. The most important region is around  1 GeV, where QCD becomes confining. How well the functions can be determined in this region will be a key concern. The vanishing momentum region is physically less relevant, although the interplay between the ghost and gluon there does determine whether or not self-consistent solutions can be obtained. Importantly, the method we use to find such self-consistent solutions is outlined in Appendix~A.

\section{The Ghost Equation}

Non-linear integral equations may admit multiple solutions and it is apparent that the ghost propagator equation falls into this class. Early attempts to include ghosts in the solution of the gluon equation used the well known infrared (IR) power law assumption for the vanishing $p^{2}$ limit of the propagators~\cite{Watson:2000,Watson:2001yv},
\begin{align}
\lim_{p^{2}\to0} \Gl(p^{2})&=a_{g}\,(p^{2})^{\kappa_{g}},\\
\lim_{p^{2}\to0} \Gh(p^{2})&=a_{c}\,(p^{2})^{\kappa_{c}},
\end{align}
the relation that is found is $\kappa_{g}=2\kappa$ and $\kappa_{c}=-\kappa$. This behavior, where the power $\kappa$ is linked, is derived from matching powers on both sides of the ghost equation using a bare ghost-gluon vertex. In solving the ghost equation the coefficients $a_{g}$ and $a_{c}$ are left unconstrained.
However  the gluon equation fixes a relation  between the two. Typically, vanishing IR gluon propagators are then considered and hence a singular ghost arises for this specific solution.

More recent studies~\cite{Boucaud:2005ce,Fischer:2008uz,Pene:2009iq,Aguilar:2009ke,RodriguezQuintero:2010wy} have found a second solution where the IR powers are not linked; a vanishing gluon admits a finite ghost solution also. There the powers are typically $\kappa_{g}=1$ and $\kappa_{c}=0$. The solution selected by physical QCD is still an open question however and at present there are ideas about the singular ghost solution being a critical endpoint of a family of solutions including the finite ones~\cite{RodriguezQuintero:2010yq}, and also suggestions of an additional gauge parameter~\cite{Maas:2009se,Cucchieri:2011um}.

It is possible that solutions in the deep IR are not solutions for the whole momentum region and this will become important when we consider the gluon equation itself. However, both of these classes of solutions can be found using numerical methods in Euclidean space using standard techniques. There is one key distinguishing feature between the two however. The infinite IR ghost solution requires the ghost to be subtracted (renormalised) at zero momentum. Its value for physically relevant momenta is then specified. How the ghost evolves is fixed. The finite solution may be found by subtraction at any value of momentum. 

We presently do not state a strong preference for either solution. However we note that only the finite solution is allowed by a perturbative renormalization scheme without fine-tuning. The infinite solution never arises without such tuning, one must search for it by specifying the infinite value at zero momentum and sacrificing such freedom in the UV. We will elaborate on this below.

\subsection{Solutions of the ghost equation alone}

\begin{figure}[tbh]
  \begin{center}
  \includegraphics[width=0.45\textwidth]{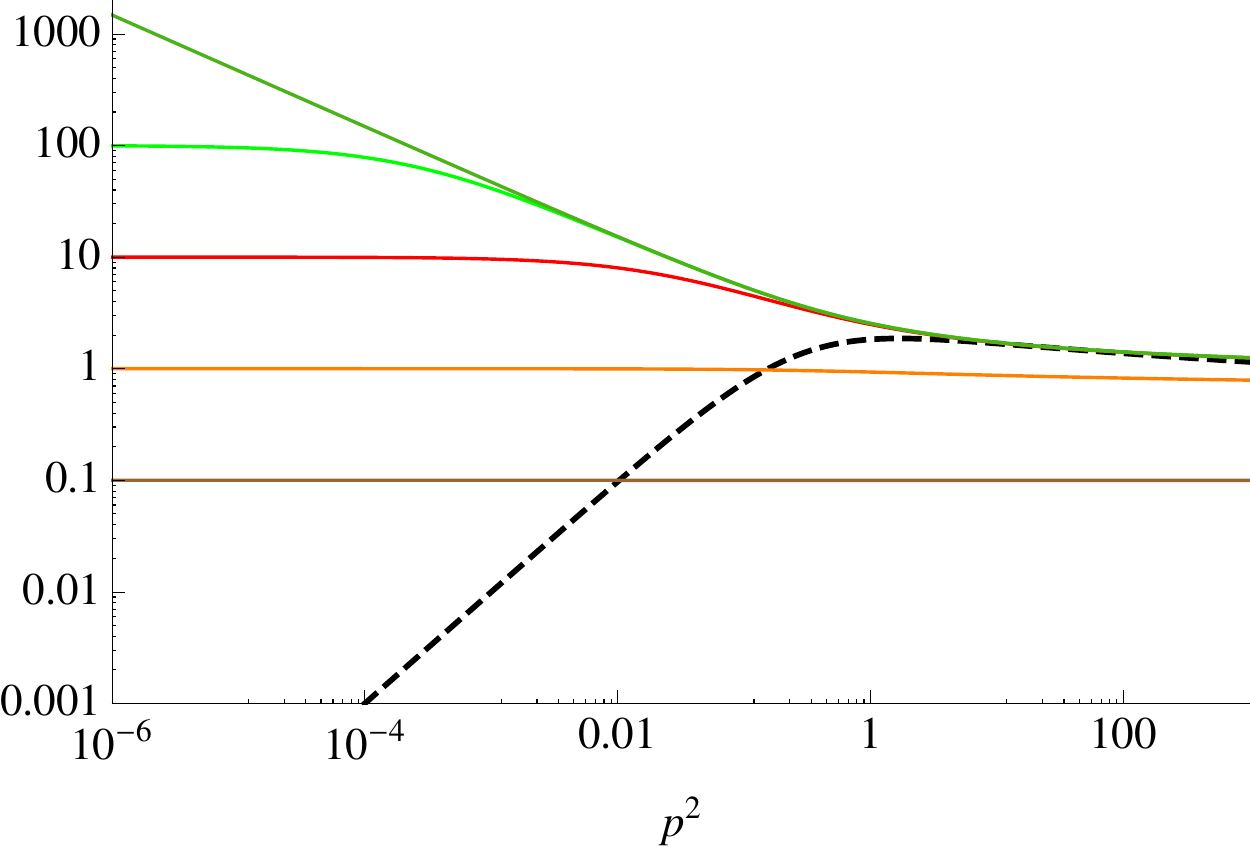}   
  \end{center}
  \caption{(Color Online) Examples of ghost solutions on a log-log plot subtracting at zero momentum. Only the specified subtraction value $\Gh(0)$ is varied between the solutions which can be read off. The $\Gh(p^{2}=0)$ value is fixed and the ghost equation solved with the depicted gluon until the ghost inputs and outputs are self-consistent. The gluon dressing $\Gl(p^{2})$, is the dashed curve and the solid curves are the different ghost dressings $\Gh(p^{2})$. The units of $p^{2}$ are arbitrary since we have not fixed the coupling to the physical value, but may be considered to be ${\cal O}\,(1$ GeV$^2$).}
  \label{fig_mixedghostsols}
\end{figure}

Using a fixed gluon input we may solve the Ghost SDE alone and investigate its sensitivity to a range of input vertices and gluons. This is useful because this type of SDE, or gap equation, is very simple to solve, and it teaches us what to expect when solving the more complicated, coupled equations self-consistently.

In doing this we utilise the bare vertex of Eq.~(\ref{eq_baregcc}), and the following model gluon~\cite{Aguilar:2010gm},
\begin{equation}
\Gl(p^{2})=\frac{p^{2}}{m^{2}+p^{2}\left(1+\frac{11}{12\pi}\frac{N_{c}g^{2}}{4\pi}\mathrm{Log}\left(\frac{p^{2}+\rho m^{2}}{\mu^{2	}}\right)\right)^{\frac{13}{22}}}
\label{eq_gluonmodel}
\end{equation}
which reproduces the correct UV perturbatively resummed logarithm, see Eq.~(B16), and provides an IR mass term necessary to produce a gluon propagator dressing function that behaves as $p^{2}$ in that limit. We utilize the parameters $\mu^{2}=10^{4}$, $m^{2}=0.1$, $g^{2}(\mu^{2})/4\pi=0.12$ and $\rho=1$ which give the gluon shown in Fig.~\ref{fig_mixedghostsols}. Dimensionful quantities can be thought of as having units close to GeV. However we do not match to a physical scale at this stage. The physical scale is determined by the value of the coupling at the renormalization point. This of course only has true physical meaning when quarks are included. This gluon, Eq.~(\ref{eq_gluonmodel}), is qualitatively similar to that found in recent Lattice QCD studies~\cite{Bogolubsky:2009dc,Cucchieri:2010xr} and in other Schwinger-Dyson studies~\cite{Fischer:2008uz,Aguilar:2009ke,Aguilar:2009nf,Aguilar:2010gm,Aguilar:2011ux}.

\begin{figure}[tbh]
  \begin{center}
  \includegraphics[width=0.45\textwidth]{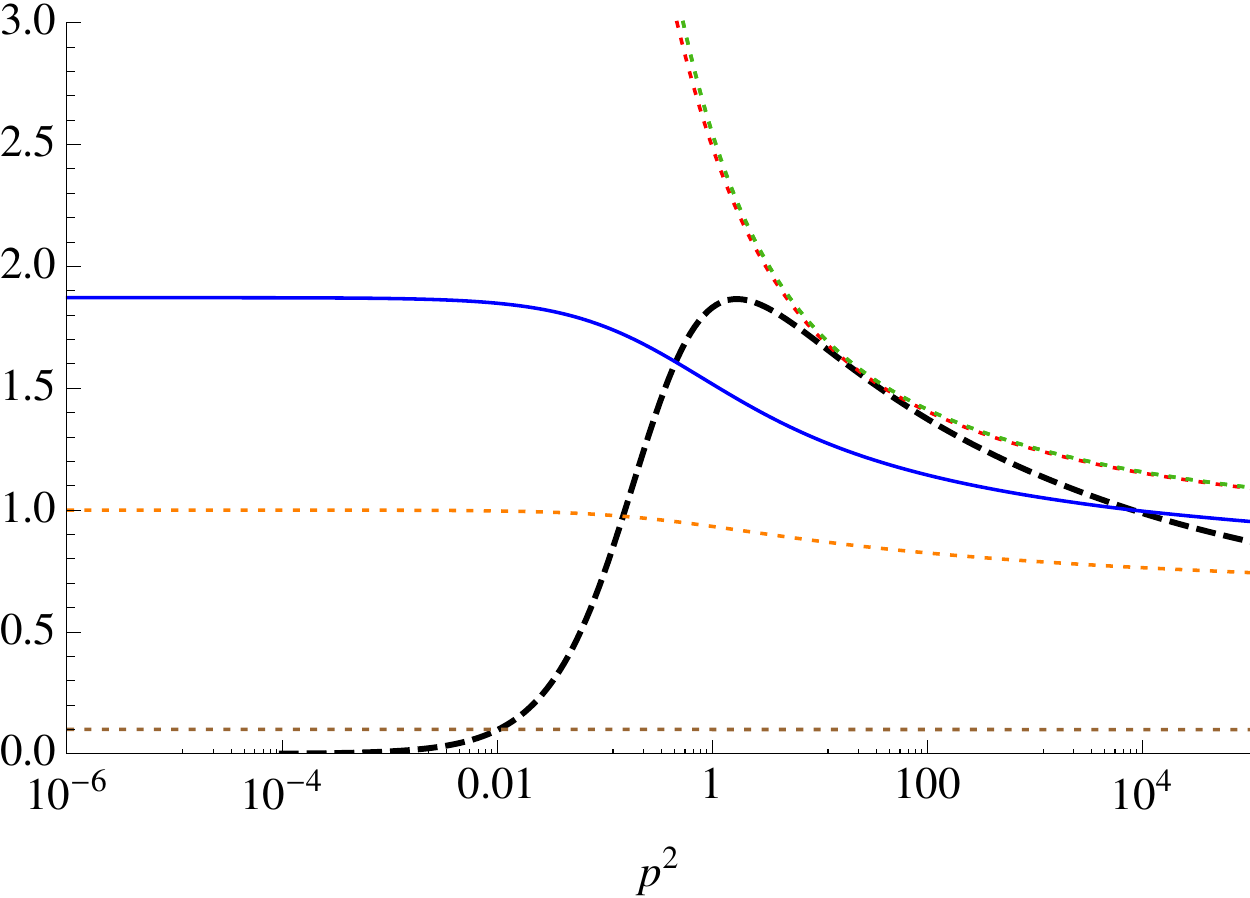}   
  \end{center}
  \caption{(Color Online) The dotted curves correspond to the zero momentum subtracted solutions and the colours match Fig.~\ref{fig_mixedghostsols}. The dashed curve is the gluon input, $\Gl(p^{2})$, and the solid curve is the physically relevant solution of the ghost equation, $\Gh(p^{2})$. The units of $p^{2}$ are arbitrary.}
  \label{fig_ghostlin}
\end{figure}

When solving the ghost equation and subtracting at zero momentum, we find its value at the subtraction point does not change its UV values between many of the solutions. Thus, for ghosts with IR values in the range $2<\Gh(0)<\infty$ we find their UV differences
to be  negligble. Note that the precise value is dependent upon the coupling and the gluon equation. For a fixed gluon, a larger coupling gives a larger critical value of $\Gh(0)$ above which all the dressings are practically indistinguishable in the perturbative region. This can be seen in Fig.~\ref{fig_mixedghostsols} where the largest three ghost dressings, although different in the IR, are identical in the UV. Below this critical value, the ghost equation admits solutions that differ in both the UV and the IR. They still exhibit the same perturbative logarithm although it is hardly visible on this scale since its coefficient is reduced when $\Gh(0)$ is set to be small. Plotted on a linear vertical axis, the effect is clearer as can be seen in Fig.~\ref{fig_ghostlin}.

This is interesting and a point that has not often been stressed. If we subtract at some perturbative value and impose the perturbative condition $\Gh(\mu^{2})=1$, and do the same for the gluon then none of the curves in Fig.~\ref{fig_mixedghostsols} are reproduced. The solution that connects to the standard perturbative solution with the standard momentum subtraction condition is unique and is separate from both the singular ghost and the zero-momentum subtraction finite solutions (except for the single fine-tuned value satified by this unique solution). 

Setting $\Gh(\mu^{2})=1$ is not essential as we can in principle renormalise at any point and the renormalization group tells us how these differently renormalised solutions are connected to each other. However, it is not possible to run down from the solution that we have obtained subtracting at a perturbative point and enforcing $\Gh(\mu^{2})=1$ to any general solution subtracted at zero momentum. Fine-tuned examples exist but these essentially predetermine the IR value.

Starting from the perturbative solution and running down into the IR, we would then stay on the finite solution for the ghost dressing for all momenta. The expectation is that asymptotically free QCD at large momentum transfers is accurately described by the perturbative solution, so this is the one we favor. In this example, a perturbative subtraction point \emph{never} admits the infinite IR ghost solution. This effect is depicted in Fig.~\ref{fig_ghostlin}, where the solid curve is the unique solution selected by renormalising $\Gh(\mu^{2})=1$ at the same $\mu^{2}$ point as the gluon. Hence, given similar conclusions from other studies~\cite{Boucaud:2005ce,Fischer:2008uz,Pene:2009iq,Aguilar:2009ke,RodriguezQuintero:2010wy,Watson:2010cn}, we carry forward this solution to investigate the gluon equation. 

In contrast, subtracting at zero momentum and specifying different values for $\Gh(0)$ maps out infinitely many other solutions even with the same  fixed gluon input. Evolving these solutions up in momentum yields different curves in the perturbative region.
We have verified that this effect is also present for fixed gluon inputs that vanish more or less rapidly in the IR, including the $\kappa\simeq 0.6$ solution~\cite{Alkofer:2003jr,Fischer:2003zc,Fischer:2008uz,Fischer:2010is}.


\subsection{Other ghost-gluon vertices in the ghost equation}

It has been noted~\cite{Boucaud:2005ce} that Taylor's theorem, which is often used to infer a bare ghost-gluon vertex, actually only places a restriction on the sum of the two functions dressing the components of the vertex, Eq.~(\ref{eq_gcctransv}). The condition is in the simplest terms, that any corrections vanish when the incoming ghost-momentum vanishes. This motivates the modification to Eq.~(\ref{eq_gcctransv}),
\begin{equation}
\Gamma_{\mu}^{(2)}=igf^{abc}\left(q_{\mu} - p_{\mu}\frac{k.q}{k^{2}}\mathcal{N}_{\mathrm{IR}}\mathcal{F}_{\mathrm{IR}}(k,p,q)\right),
\label{eq_gcctt}
\end{equation}
which is clearly not transverse to the gluon momentum. However, the piece additional to the bare term now vanishes when the incoming ghost momentum $p$ vanishes. The non-perturbative normalization parameter $\mathcal{N}_{\mathrm{IR}}$ is also introduced into the modeling. Unlike dressing forms that add only to the $k_{\mu}$ term, this form may affect the solutions of the ghost equation, as we demonstrate below in Fig.~\ref{fig_ghosttay}. Adopting the same method as above, we fix the gluon using the previous model and parameters, and solve the ghost until self-consistent with this input and its own equation.

\begin{figure}[tbh]
  \begin{center}
  \includegraphics[width=0.45\textwidth]{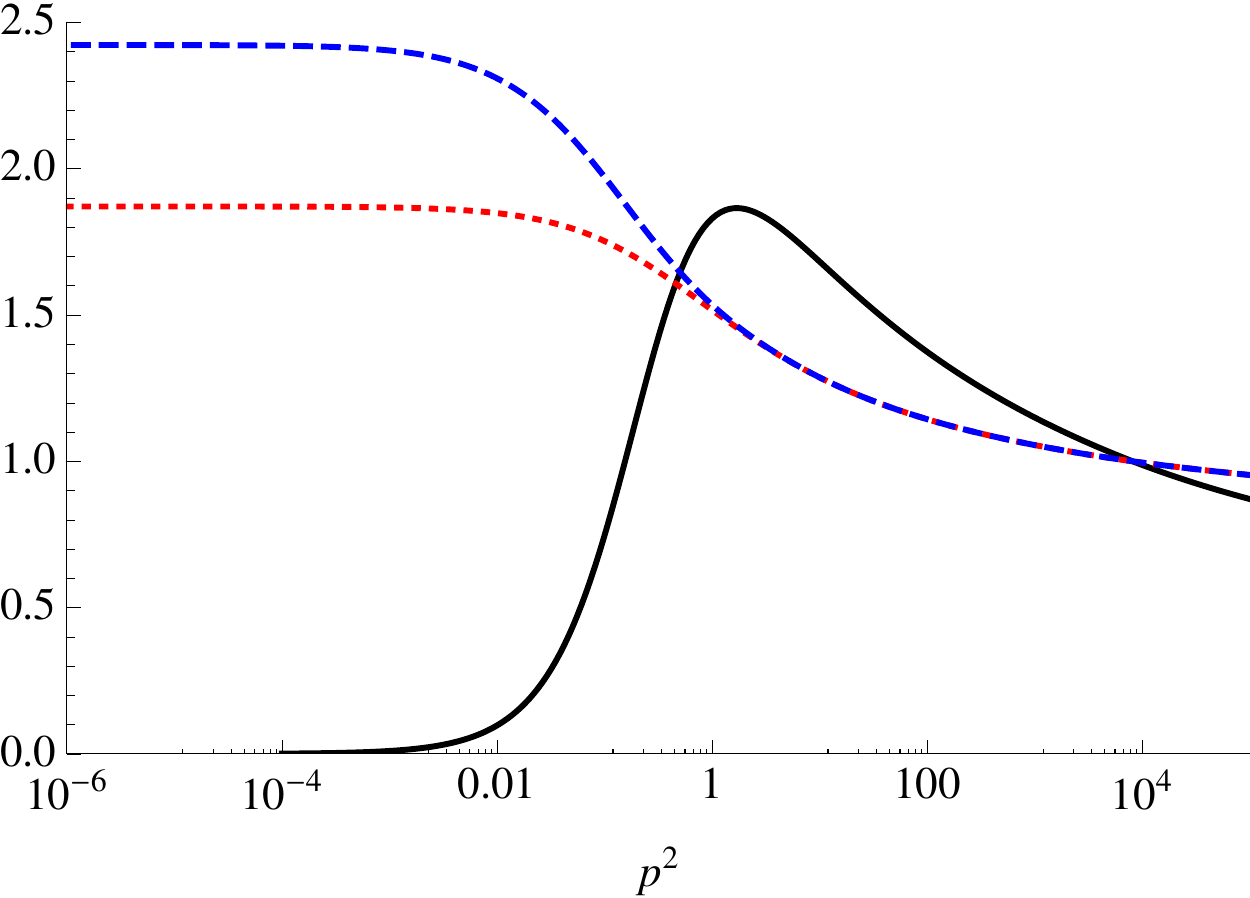}   
  \end{center}
  \caption{(Color Online) A comparison of two solutions normalised using the perturbative condition. The (red) dotted curve is the ghost solution, $Gh(p^{2})$, with a bare vertex or transverse vertex. The (blue) dashed curve is the ghost dressing obtained using Eq.~(\ref{eq_gcctt}). The (black) solid curve is the fixed gluon dressing input, $\Gl(p^{2})$ from Eq.~(\ref{eq_gluonmodel}). The units of $p^{2}$ are arbitrary internal units.}
  \label{fig_ghosttay}
\end{figure}

We find that in principle the ghost equation is dependent upon the choice of vertex and fairly large corrections are possible as found for the example shown in Fig.~7. Many other extensions are possible and these exist in the literature. Clearly a precise form for this vertex would be useful. However at present we must consider this part of the uncertainty on the result of the calculation. Corrections of 20\% over the bare vertex appear to be allowed. Nevertheless we have found the ghost equation to be solvable for a wide range of parameters and choices of vertices.

\section{Coupling Gluons and Ghosts}

After the ease of solving the ghost equation, we now turn to the gluon, where matters are very different.

It is the opinion of the present authors that no completely satisfactory solution exists for the gluon propagator equation at present. Solutions exist, but they are highly dependent on arbitrary vertex choices, and disappointingly this is an issue rarely addressed. The predictive power of the equations is often lost through the introduction of arbitrary parameters that are not derived from the fundamental field theory. The simplest possible model just modifies the gluon propagator, such that it contains an IR mass term that vanishes in the UV. Modeling vertices with a number of parameters and a range of possible forms is unfortunately no more predictive. However, it can be useful in guiding where to look and in understanding those quantities we may need to know precisely and those that are less important.

Although we intend to subtract and renormalise at a perturbative point, which excludes the singular ghost solution, it is worth saying a few words about one particular singular solution, especially in the light of recent criticism~\cite{Fischer:2010is}. In what follows, we note that this is not directed at all singular solutions, just this particular one and any others that contain the same flaw we describe.

The problem arises due to the combined effects of truncation and projection of the dressed gluon propagator in the Landau gauge,
\begin{align}
\mathcal{D}_{\mu\nu}(p)&=\frac{\Gl(p^{2})}{p^{2}}\left(g_{\mu\nu}-\frac{p_{\mu}p_{\nu}}{p^{2}}\right),\\
&=\mathcal{A}(p^{2})g_{\mu\nu}-\mathcal{B}(p^{2})\frac{p_{\mu}p_{\nu}}{p^{2}}.
\end{align}
The structure of the gauge theory demands that $\mathcal{A}(p^{2})=\Gl(p^{2})=\mathcal{B}(p^{2})$, which in a full treatment with no truncation would be the case. However, in an incomplete treatment where uncontrolled approximations are made great care must be taken, because individual diagrams contain quadratic divergences that are no longer guaranteed to cancel.

The simplest solution has been long known and is straightforward~\cite{Brown:1988bm,Brown:1988bn,Bloch:1994if}. In solving the gluon equation we only need to determine $\mathcal{A}$ or $\mathcal{B}$. As is known from perturbation theory, the problematic quadratic divergences only occur in the $\mathcal{A}$ term, so we just project onto $\mathcal{D}_{\mu\nu}$ of Eqs.~(15,16) in such a way that only the $\mathcal{B}$ term is retained. This is achieved using the so-called \emph{Brown-Pennington} projector~\cite{Pennington:2010gy},
\begin{equation}
\mathcal{P}_{\mu\nu}(p)=g_{\mu\nu}-d\,\frac{p_{\mu}p_{\nu}}{p^{2}}
\end{equation}
where $d$ is the number of dimensions in which we work. Using the $\mathcal{A}$ term leaves any solution exposed to unphysical quadratic divergences, and these are present in one set of solutions in the literature~\cite{Alkofer:2003jr,Fischer:2010is}. The solutions are obtained by the insertion of an additional step in the iterative procedure, where the quadratic divergence is subtracted from the \emph{result} of the gluon loop in the gluon equation. This is not necessarily a safe thing to do since in the non-perturbative region the integration results are \emph{a priori} unknown. This point has been made previously~\cite{Bloch:2003yu} and is of fundamental importance. 

\begin{figure}[tbh]
  \begin{center}
  \includegraphics[width=0.45\textwidth]{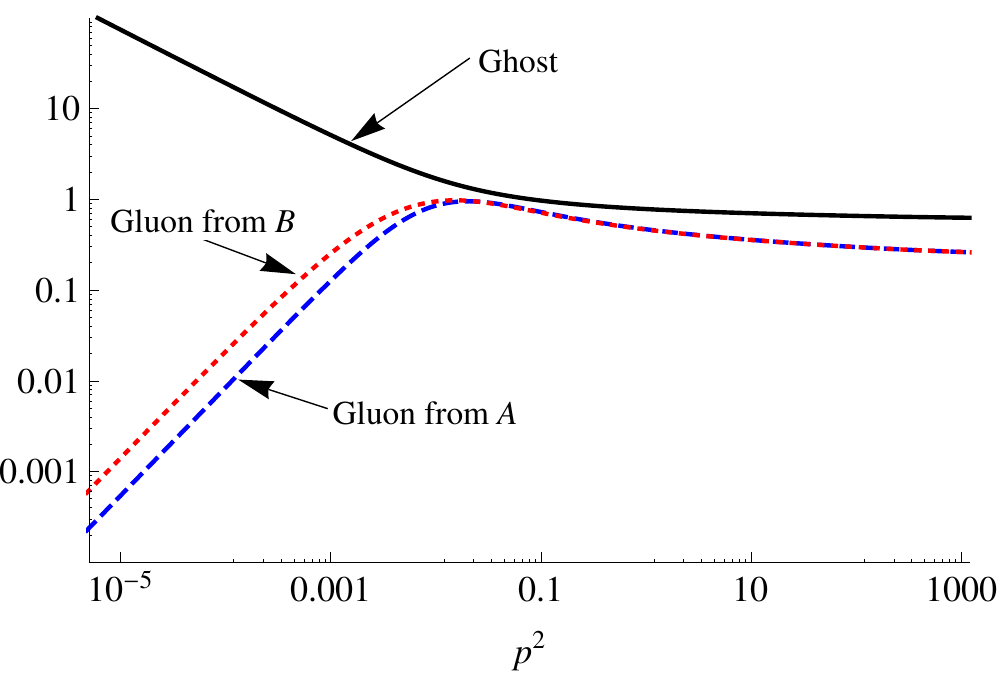}   
  \end{center}
  \caption{(Color Online) The singular ghost solutions and violation of transverality in the infrared. The dashed gluon and dotted ghost curves are obtained first using the $\mathcal{A}$ term and then we switch to the $\mathcal{B}$ term and we find that transversality is broken since the curves differ. The units of $p^{2}$ are arbitrary internal units.}
  \label{fig_scaling}
\end{figure}

In the IR analysis which is common to this solution and all subsequent refinements, the gluon determined from $\mathcal{B}$ differs from the gluon determined from $\mathcal{A}$ signaling the breaking of transversality of the propagator. This may be seen in~\cite{Bloch:2003yu} in the IR analysis, where the $\mathcal{A}$ and $\mathcal{B}$ terms are analysed separately and the difference is clear. It may also be seen by solving self-consistently for the singular solution defined by the $\mathcal{A}$ function, and then seeing how the solutions change when we switch from the $\mathcal{A}$ term and the $\mathcal{B}$ term. This we show in Fig.~\ref{fig_scaling} and we see a clear difference between the two gluon dressings, signalling there is an issue with transversality. Moreover the $\mathcal{B}$ solution is not self-consistent. If the usual iterative procedure is followed then no self-consistent solutions can be found where the $\mathcal{B}$ term satisfies the set of equations~\footnote{We do note however that other infinite ghost solutions exist and that the functional renormalization group methods also find this solution. We do not comment on these further other than as stated above: we do not attempt to rule out any other infinite ghost solutions.}.

We thus determine our gluon from the $p_{\mu}p_{\nu}$ term of the propagator which is known to yield the correct logarithms in the UV limit.
Reference~\cite{Fischer:2008uz} addresses this issue by introducing parameters into their vertices to make ${\mathcal{A}}$ and ${\mathcal{B}}$ come together.

\subsection{Self-consistent solutions of both propagators in a one-loop only system}

We proceed in the traditional manner by starting with the simplest conceivable system, then attempt to obtain solutions. In this system, we neglect the two two-loop dressed contributions in Fig.~1 to the gluon propagator and also drop the quark interaction for now. The required input is the triple-gluon vertex and the ingredients described above for the ghost equation.

There are several sources of information regarding the triple-gluon vertex. One property that we consider important that tends to be neglected is the Bose symmetry of the vertex. Bose symmetry is, of course, present at all orders in perturbation theory, and in the solution of the WSTI for this vertex~\cite{Davydychev:1996pb,vonSmekal:1997vx}.

We proceed by considering three dressings of the triple-gluon vertex of increasing complexity, we factor off $igf^{abc}$, and write the bare vertex Lorentz structure as $\Gamma^{(0)}_{\mu\nu\rho}(k,p,q)$,
\begin{align}
\Gamma^{(A)}_{\mu\nu\rho}(k,p,q)&=\Gamma^{(0)}_{\mu\nu\rho}(k,p,q)\frac{\Gh(p^{2})\,\Gh(q^{2})}{\Gl(p^{2})\,\Gl(q^{2})}\\
\Gamma^{(B)}_{\mu\nu\rho}(k,p,q)&=\Gamma^{(0)}_{\mu\nu\rho}(k,p,q)\times\nonumber\\
\frac{1}{3}&\left(\frac{\Gh(k^{2})}{\Gl(k^{2})}+\frac{\Gh(p^{2})}{\Gl(p^{2})}+\frac{\Gh(q^{2})}{\Gl(q^{2})}\right)\\
\Gamma^{(C)}_{\mu\nu\rho}(k,p,q)&=\frac{1}{2}\left(\frac{\Gh(q^{2})}{\Gl(p^{2})}+\frac{\Gh(q^{2})}{\Gl(k^{2})}\right)g_{\mu\nu}(k-p)_{\rho}\nonumber\\
                                &+\frac{1}{2}\left(\frac{\Gh(k^{2})}{\Gl(q^{2})}+\frac{\Gh(k^{2})}{\Gl(p^{2})}\right)g_{\nu\rho}(p-q)_{\mu}\nonumber\\
                                &+\frac{1}{2}\left(\frac{\Gh(p^{2})}{\Gl(k^{2})}+\frac{\Gh(p^{2})}{\Gl(q^{2})}\right)g_{\rho\mu}(q-k)_{\nu}\; .
\label{eq_gggwsti}
\end{align}
The first vertex, $\Gamma^{({A})}$, has been used to reproduce the precise perturbatively resummed one-loop running of the gluon propagator~\cite{Bloch:2001wz}, note that it does not reproduce the one-loop behavior of the triple-gluon vertex itself, nor does it have Bose-symmetry between each leg. The second vertex, $\Gamma^{({B})}$ is a simple symmetric vertex inspired by WSTI solutions that involve ratios of dressing functions. Finally, $\Gamma^{({C})}$ is an approximate solution to the WSTI itself using a bare ghost-gluon scattering kernel ($\tilde{\Gamma}_{\mu\nu}=g_{\mu\nu}$). Since the ghost-gluon vertex WSTI and hence the ghost-gluon scattering kernel contributions are not precisely known, we do not go further than this at present. A full solution of the triple-gluon WSTI using an approximate ghost-gluon scattering kernel is available in the literature~\cite{Ball:1980ay,Ball:1980ax,vonSmekal:1997vx}, however the associated ghost-gluon vertex is insufficient to obtain solutions in a finite ghost system.

We select a conservative set of parameters that from experience allow solutions for our vertices given above. These are not intended to describe the physical world in any way, but rather to expose the salient features of the truncation. For sensible comparison, we only use the same parameters, where applicable.

\begin{figure}[tbh]
  \begin{center}
  \includegraphics[width=0.45\textwidth]{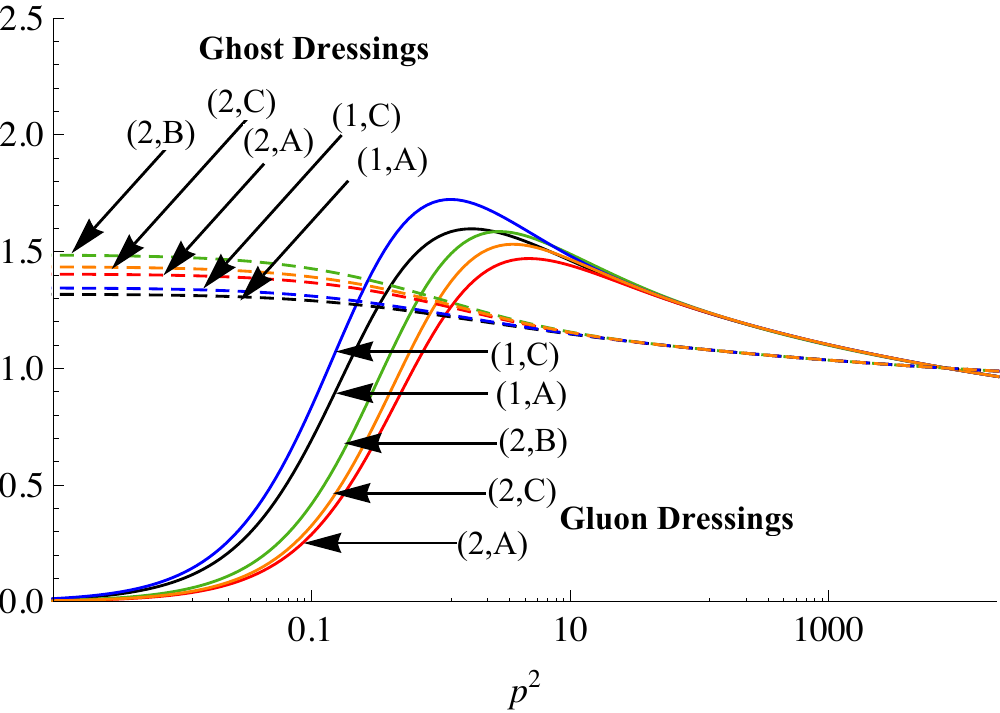}   
  \end{center}
  \caption{(Color Online) The range of solutions obtained using the possible vertex combinations. The label $(i,j)$ refers to the vertices used in obtaining the solutions corresponding to $\Gamma^{(i)}_{\mu}$ for the ghost-gluon vertex and $\Gamma^{(j)}_{\mu\nu\rho}$ for the triple-gluon vertex. The missing curve corresponds to the $\Gamma^{(1)}_{\mu}$ and $\Gamma^{(B)}_{\mu\nu\rho}$, self-consistent solutions were not obtained there. The parameters are not varied between the solutions, only the vertices. The units of $p^{2}$ are arbitrary since we have not fixed the coupling to a physical value.}
  \label{fig_vertices}
\end{figure}

The solutions obtained are shown in Fig.~\ref{fig_vertices}. We label the solutions using $(i,j)$ to denote the vertices we use. These are defined above and correspond to $\Gamma^{(i)}_{\mu}$ for the ghost-gluon vertex and $\Gamma^{(j)}_{\mu\nu\rho}$ for the triple-gluon vertex. Similarly to the solution of the ghost equation alone, we find a strong sensitivity to the vertices we choose, particuarly in the all-important region that is most relevant to physics: roughly between 0.1 and 10 in these momentum-squared units (see Figs.~(\ref{fig_initialplot}-\ref{fig_rc})). Although we do not match to a specific scale, the peak of the gluon dressing function would be expected to have a close relation to the fundamental scale $\Lambda_{\mathrm{QCD}}$. The differences between these solutions with the range of vertices $\{i,j\}$ would undoubtedly have an effect on physical quantities, for example, hadron masses and form-factors~\cite{Maris:1997hd}.

\subsection{Infrared Loop Contributions from different vertices}

\begin{figure*}[!hbt]
  \begin{center}
  \includegraphics[width=0.45\textwidth]{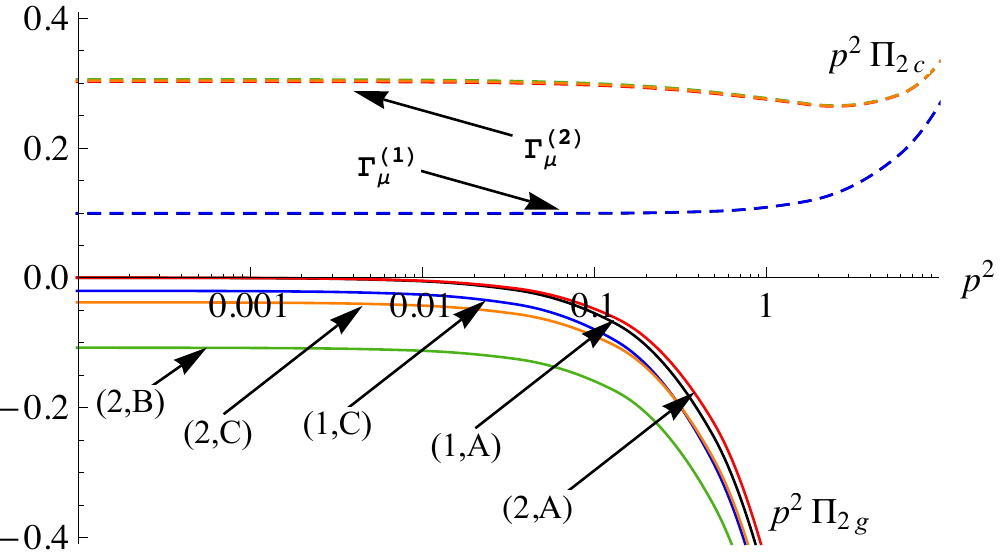}  
  \includegraphics[width=0.45\textwidth]{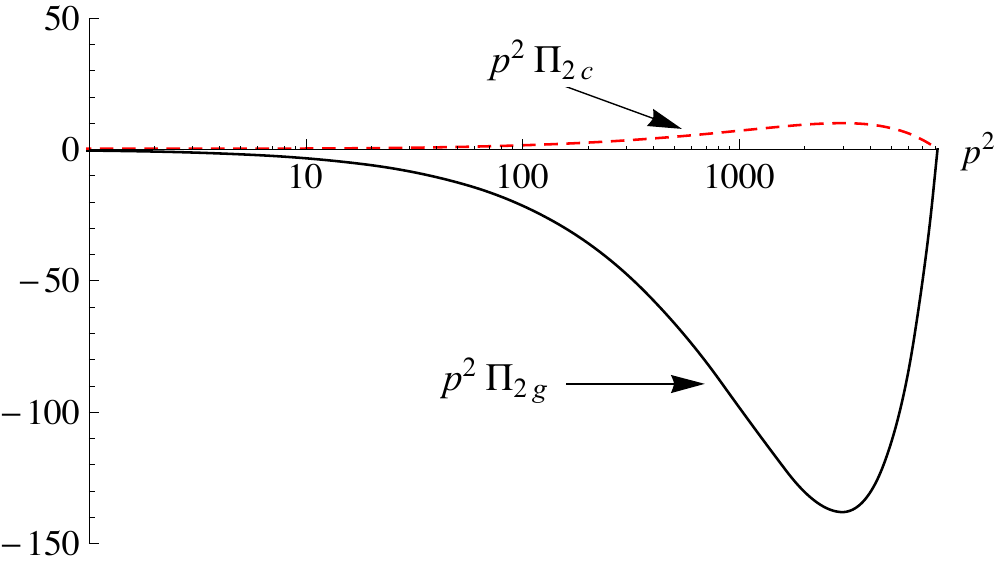}    
  \end{center}
  \caption{(Color Online) Gluon polarization functions, $p^{2}\Pi_{2c}(p^{2})$ (positive values) and $p^{2}\Pi_{2g}(p^{2})$ (negative values). Left: IR region, Right: UV region. In the UV the curves are indistinguishable on this scale. In the IR, the ghost-loop curves for multiple solutions lie on top of each other. The bare ghost-gluon vertex, not shown here, always gives a vanishing IR contribution given an IR vanishing gluon dressing and an IR finite ghost dressing. The functions $p^{2}\Pi_{2c}(p^{2},\mu^{2})$ are labelled according to the input vertices with the contributions from $\Gamma_{\mu}^{(i)}$ and $\Gamma_{\mu\nu\rho}^{(j)}$ labelled as $(i,j)$ in the plot. The units of $p^{2}$ are arbitrary.
  }
  \label{fig_pols}
\end{figure*}


We note that previous studies have found ghost-dominance in the IR region of the gluon equation. We show this to be false and an artifact of choosing a simple vertex. We find that as in the perturbative region, the ghost and gluon loops contribute a similar numerical amount. This may negate arguments relating to a ghost-gluon vertex that is transverse alone. 

In order to explain this, we write the renormalised (subtracted) gluon equation in the following form,
\begin{equation}
\Gl(p^{2})^{-1}=\Gl(\mu^{2})^{-1}+\Pi_{2g}(p^{2},\mu^{2})+\Pi_{2c}(p^{2},\mu^{2}),
\end{equation}
the functions $\Pi(p^{2},\mu^{2})$ are the loop integrals subtracted at the renormalization point $\mu^{2}$. These polarization functions are then zero at $p^{2}=\mu^{2}$ and $\Gl(\mu^{2})=1$. It is expected that the gluon propagator dressing should be positive for all momenta.

The ghost-loop polarization function, $\Pi_{2c}(p^{2},\mu^{2})$, is highly sensitive to the ghost-gluon vertex, and the gluon loop polarization function, $\Pi_{2g}(p^{2})$, is most sensitive to the choice of triple-gluon vertex. This is completely as expected. There are additional feedback effects that happen when the equations are solved together as described in appendix \ref{app:num}, but these effects are typically smaller. These functions diverge as $p^{2}$ (or a little less) so we multiply up by $p^{2}$ and plot the numerical results in Fig.~\ref{fig_pols}.

In Fig.~\ref{fig_pols} we see the key feature that we wish to elucidate, and that is, for the Bose-symmetric triple-gluon vertices, the IR gluon-loop contributions are non-zero. Both ghost and gluon loops contribute at a similar numerical order, whilst the ghost must be larger to keep the gluon dressing function positive in the IR limit. There is clearly some interplay between these, as can be seen by the differences between the transverse vertex and the modified version that satisfies Taylor's theorem. An interdependence of this type is to be expected, since the ghost-gluon scattering kernel appears separately in the WSTI for both vertices, and the idealised full solution would uniquely constrain this relation.

The behavior of these functions is important, particularly in light of the above partial cancellation, if the gluon propagator dressing function is to be positive everywhere. As the loop contribution induced by the triple-gluon dressing is negative then some canceling positive contribution is required from the loop containing the ghost-gluon vertex. This is only present for the two vertices we show here. For the bare vertex or a vertex dressed by simple ratios, no such contribution occurs and hence self-consistent solutions are not possible.
It is this extreme sensitivity, highlighted in Fig.~10, to the ans\"atze for both the ghost-gluon and triple gluon vertices that makes us conclude that consistent gluon and ghost propagators have not yet been determined in continuum strong coupling QCD.

An alternative to this would be for a positive contribution to arise from the gluon loop. This would require further corrections to be added to the triple-gluon vertex. Some evidence for this may already exist in recent lattice studies \cite{Cucchieri:2008qm}. These contributions could be due to the WSTI-unconstrained parts of the triple-gluon vertex and hence are not present in $\Gamma^{(C)}_{\mu\nu\rho}$ of Eq.~(\ref{eq_gggwsti}), or a more complete WSTI solution~\cite{vonSmekal:1997vx}.

A point of note is that the coefficient of $p^{2}$ in the IR limit of the gluon propagator dressing is determined by the sum of the ghost and gluon loop functions in that limit. In order to determine this with any precision it is important to get the vertices right. This coefficient is of some importance since it determines the effective gluon `mass' term. The mass term, though almost certainly gauge dependent, will inevitably affect physical quantities like hadron masses and form-factors. Many practical models contain mass terms for the gluon; for a recent example see~\cite{Qin:2011dd}. 

Considering these two contributions from the different loops we also find the reason why the gluon dressing looks similar in both the infinite and finite ghost solutions. In the singular solution the bare vertex is used and the large contribution from the ghost loop is provided by the singular propagator dressing. Conversely in the finite ghost solution a non-trivial vertex provides a similar large contribution and a qualitatively similar vanishing gluon arises.

\section{Comparison to Lattice QCD}
Primarily we have been motivated by theoretical issues encountered in solving the Schwinger-Dyson equations for the gluon and ghost propagators. However a complementary technique is available where these quantities have been calculated and that is Lattice QCD. The possible issues there are quite different to those that we may induce here by truncation, so a comparison is a useful independent cross-check. Importantly the lattice computations, for which there are extremely precise results, are for the pure gauge sector we investigate here in the continuum.

Many recent lattice studies exist~\cite{Cucchieri:2007md,Bicudo:2010wi,Cucchieri:2011ga}, starting with the early work of~\cite{Mandula:1987rh}. We compare our calculations to \cite{Bogolubsky:2009dc} which provides results in Landau gauge for both dressing functions. The qualitative behavior there is as we have here with a finite ghost dressing function and a finite gluon propagator, corresponding to a vanishing gluon propagator dressing function.	

\begin{figure*}[thb]
  \begin{center}
  \includegraphics[width=0.45\textwidth]{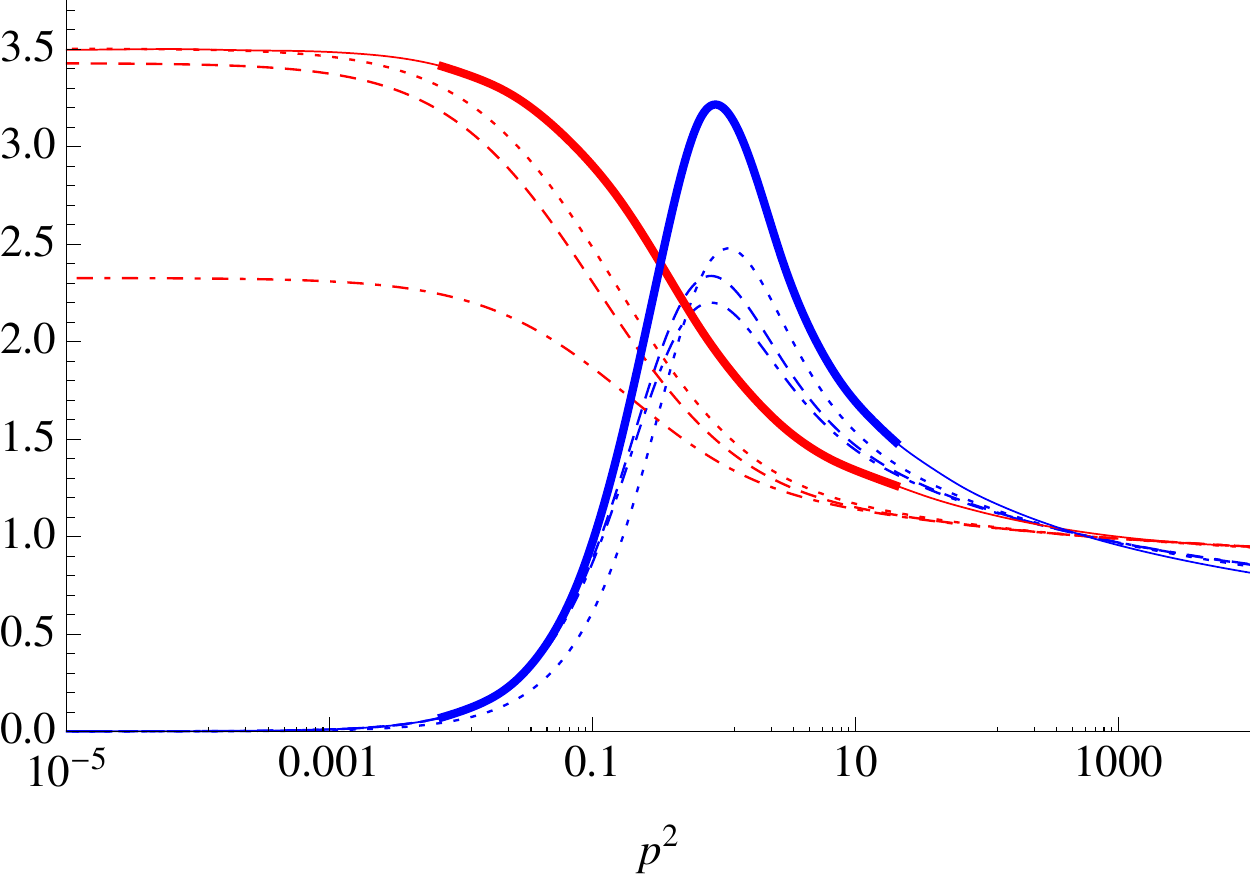}   
  \includegraphics[width=0.45\textwidth]{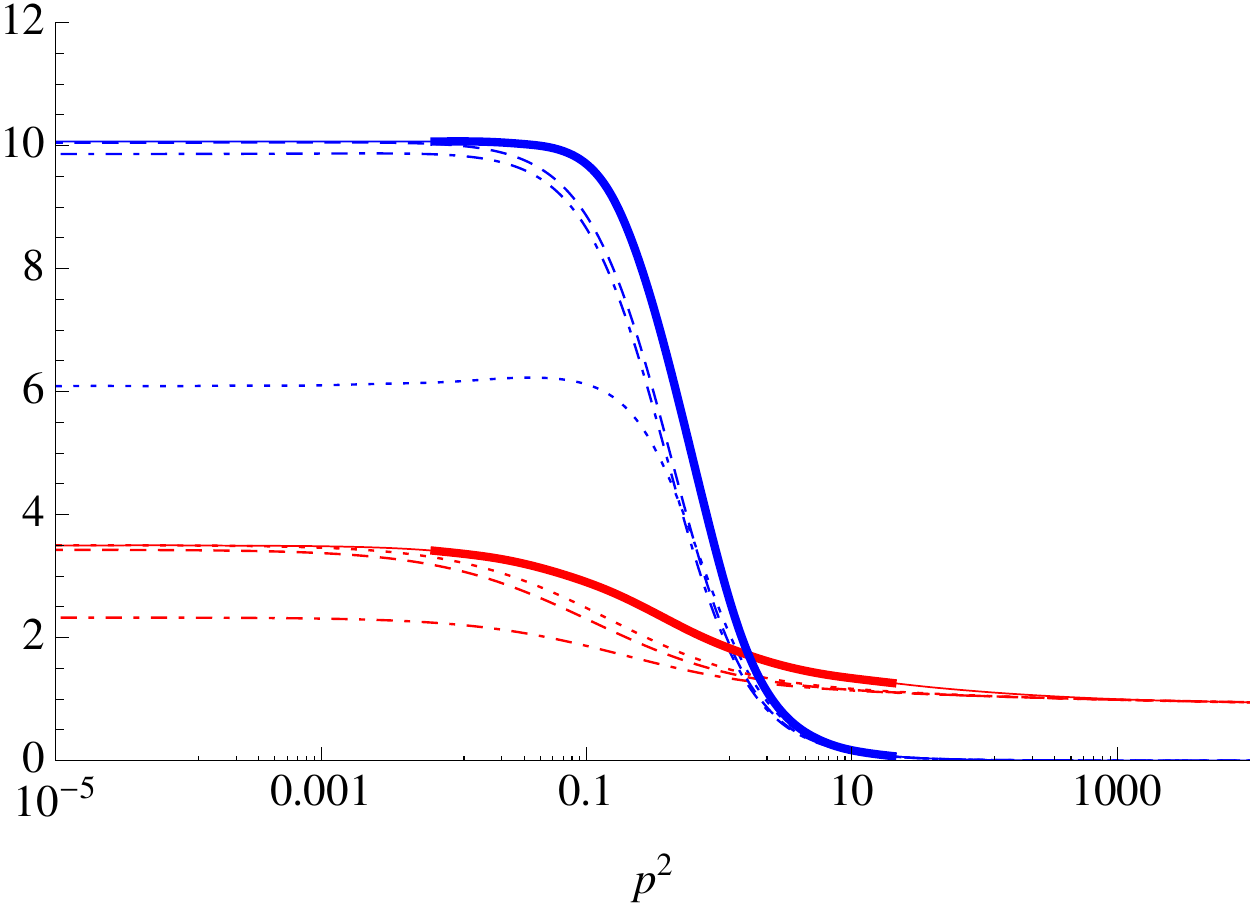}
  \end{center}
  \caption{(Color Online) Self-consistent solutions tuned to lattice solutions, showing the dressing functions. The solid curves depict a smooth fit to the lattice data. The heavier region is where the functions are represented by lattice data and the feint region represents the natural extrapolation. The broken curves are the tuned solutions. Left: The blue peaked curves correspond to the gluon dressing function $\Gl(p^{2})$, this vanishes as $p^{2}\to0$, the red monotonic curves correspond to the ghost dressing function $\Gh(p^{2})$. Right: The upper (blue) curves correspond to $\Gl(p^{2})/p^{2}$, this is $\sim10$ as $p^{2}\to0$, the lower (red) curves correspond to the ghost dressing function.}
  \label{fig_latdf}
\end{figure*}

In obtaining these solutions we use the preferred set of vertices, $\Gamma^{(C)}$ from Eq.~(\ref{eq_gggwsti}) and the ghost-gluon vertex given in Eq.~(\ref{eq_gcctt}). We then tune the parameters to obtain a reasonable representation of the lattice data and these are given in Table~\ref{tab_latticepars}. We note several numerical discrepancies. Most importantly in the UV, the available parameter space does not include the coupling strength found on the lattice; smaller couplings have to be used. The iterative procedure  described in Appendix~A breaks down at larger couplings signalling an absence of solutions in this given truncation. This is visible in the plots as a smaller gradient of the perturbatively resummed log for the SDE solutions. Secondly, the peak of the gluon dressing function in Fig.~\ref{fig_latdf} is not as large as found on the lattice. 
It is expected that the two-loop dressings would make a contribution here, so that could be the source of the difference. Other effects will of course include 
the different coupling values and uncertainties in the vertices. We do not comment upon any differences due to the method used to extract the predictions from the lattice.

\begin{table}[htb]
\begin{center}
\begin{tabular}{c|cccc}
\hline
Solution & $\;\mu^{2}\;$ & $\;\alpha(\mu^{2})\;$ & $\;\Lambda^{2}_\mathrm{IR}\;$ & $\;\mathcal{N}_\mathrm{IR}\;$\\
\hline
Dotted~ & ~~~650~~ & ~~0.1313~~ & ~~6.0~~ & ~~3.5~~ \\
Dot-dashed~ & ~~~650~~ & ~~0.1200~~ & ~~3.8~~ & ~~3.7~~ \\
Dashed~ & ~~~650~~ & ~~0.1225~~ & ~~7.5~~ & ~~2.0~~ \\
\hline
\hline
\end{tabular}
\end{center}
\caption{Parameters used in obtaining the tuned self-consistent lattice solutions. $\mu$ and $\Lambda_\mathrm{IR}$ can be regarded as being in GeV units.}
\label{tab_latticepars}
\end{table}

\section{Outlook}

The gluon and ghost propagators are the basic Green's functions that embody not only the short distance behavior determined by the asymptotically free nature of QCD, but confinement dynamics at larger distances. This behavior is encoded in solutions of the appropriate Schwinger-Dyson equations. 
Here we first investigated the existence of solutions for a simplified ghost equation using a fixed gluon input. Multiple solutions were found. However applying a condition commonly used in perturbative analyses, that $\Gh(\mu^{2})=1$ where $\mu^{2}$ is the point in the perturbative region at which both gluon and ghost are renormalised, we found just one solution was preferred. Using this result we then investigated the necessary terms required in order to construct a self-consistent solution for the coupled gluon and ghost dressing functions with input interaction vertices. A range of solutions resulted that are qualitatively similar to those computed on the lattice. However, perfect quantitative agreement has yet to be established. This is assumed to be due to the approximations made here or on the lattice. 
We drew special attention to differences induced by vertex choices and we have argued that the neglected 
two-loop dressings are likely to give rise to significant changes. The hope is that constraints on the vertices, particularly that of the full ghost-gluon interaction, can be found that uniquely specify their structure. At present the \emph{ad hoc} vertices used are motivated more by the practicalities of finding solutions than by constraints derived from the fundamental field theory.
Thus we conclude that consistent gluon and ghost propagators have yet to be determined in continuum strong coupling QCD.

\begin{figure}[tbh]
  \begin{center}
  \includegraphics[width=0.45\textwidth]{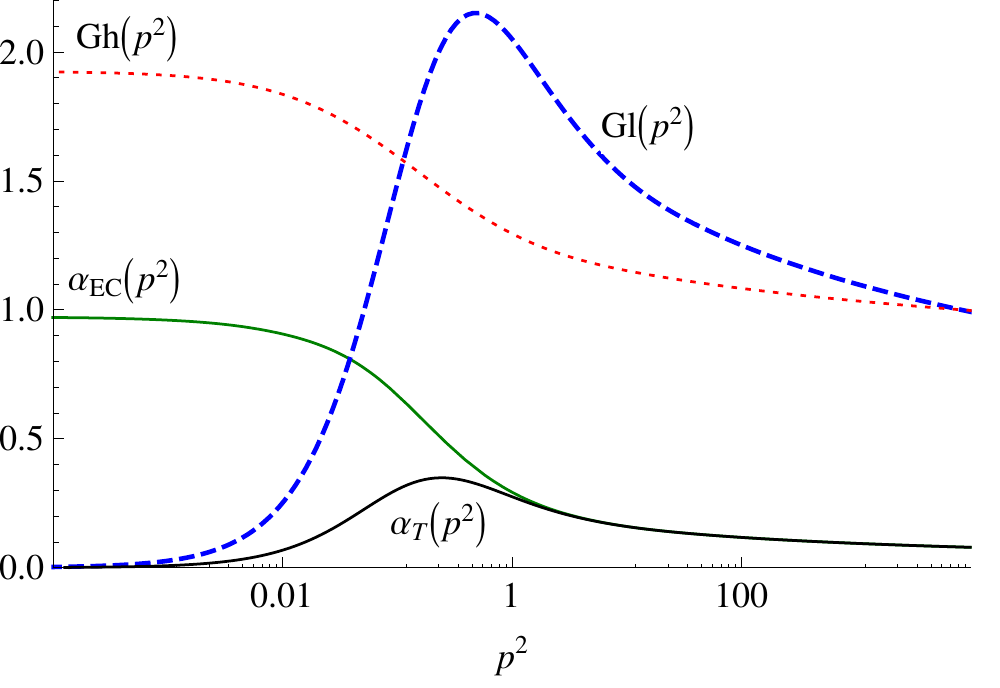}   
  \end{center}
  \caption{(Color Online) The running coupling $\alpha_{T}(p^{2})$ from Eq.~(\ref{eq_rctaylor}) is shown as the black solid curve. The running coupling $\alpha_{EC}(p^{2})$ from Eq.~(\ref{eq_alphaec}) is shown as the green solid curve. The gluon dressing $\Gl(p^{2})$ is the dashed blue curve and the dotted red curve is the ghost dressing function $\Gh(p^{2})$. The units of $p^{2}$ are now very close to GeV$^{2}$.}
  \label{fig_rc}
\end{figure}

In the pure gauge sector studied here, one can define a non-perturbative running coupling following Taylor~\cite{Taylor:1971ff} from the ghost-gluon vertex renormalization, 
\begin{equation}
\alpha_{T}(p^{2})=\frac{g^{2}(\mu^{2})}{4\pi}\;\Gh^{2}(p^{2})\,\Gl(p^{2})
\label{eq_rctaylor}
\end{equation}
where $\mu^{2}$ is the renormalization point and $g(\mu^{2})$ fixes the physical scale. Experiment with 5 flavors of fermion gives $\alpha(M_\mathrm{Z}^2)\,=\,0.118$ in the modified minimal subtraction ($\overline{MS}$) scheme. This can be related to momentum subtraction with zero flavors appropriate to the calculations we performed here, by using perturbative results from the two schemes~\cite{Celmaster:1979km,Chetyrkin:2000fd,Gracey:2011pf}. Thus, this value of $\alpha$ in the $\overline{MS}$ scheme with 5 flavors translates into $\alpha(M_Z^2)\,=\,0.08$ with zero flavors in the momentum subtraction scheme. Then with $\Lambda_{\mathrm{IR}}=2$ GeV and $\mathcal{N}_{\mathrm{IR}}=2$ we obtain the function $\alpha_{T}(p^{2})$ shown in Fig.~\ref{fig_rc}.

The behavior of the gluon propagator dressings in Figs.~\ref{fig_initialplot}, \ref{fig_vertices}, \ref{fig_latdf} and \ref{fig_rc} can be interpreted in terms of the dynamical generation of an effective gluon mass. This in turn leads to the following proposal for a corresponding definition of the running coupling
\begin{align}
\alpha_{EC}(p^{2})=\frac{g^{2}(\mu^{2})}{4\pi}\;\frac{p^2+m_g^{2}(p^2)}{p^2}\;\Gh^{2}(p^{2})\Gl(p^{2})\,,
\label{eq_alphaec}
\end{align}
recently developed in a study of the effective charges given by dressing functions with such qualitative behavior~\cite{Aguilar:2009nf}. Applying this to the solutions found using $\alpha(M_Z^2)\,=\,0.08$, $\Lambda_{\mathrm{IR}}=2$ GeV and $\mathcal{N}_{\mathrm{IR}}=2$ we find $m_g(0)=350$ MeV and the function $\alpha_{EC}(p^{2})$ in Fig.~\ref{fig_rc}. The result is not particularly sensitive to the precise form of the function $m_{g}(p^{2})$, however we use that given in~\cite{Aguilar:2009nf} which is suppressed in the UV. 

Both gluon and ghost dressings do not rise as steeply as the lattice-like solutions of Fig.~11.  Nevertheless, the collective effect of the gluon dressing and this running coupling would just have sufficient strength to induce the dynamical generation of quark mass, {\it cf.} the model of Maris and Tandy~\cite{Maris:1997hd,Maris:1999nt}. The inclusion of quarks is, of course, an important aspect of making connection to physical quantities, like the pion mass and its decay constant. A complete study requires a simultaneous solution of the quark SDE with those of the gluon and ghost. Research over the past 15 years indicates that we know what the quark propagator functions look like for a whole range of quark masses (and hence flavors) with model gluon inputs.
Indeed modeling such ghost and gluon propagators and their interactions, and inserting these into the Bethe-Salpeter equations to compute hadronic observables is now well advanced~\cite{Maris:1997hd,Maris:1999nt,Chang:2009zb,Qin:2011dd}. However, a comprehensive computation with a self-consistent investigation of  the coupled quark, gluon and ghost equations in continuum QCD,  followed by the development of a trustworthy hadron phenomenology, is still in the future. The issues exposed by the present study show we have some way to go before we can claim robust results from such an {\it ab initio} approach.

\section*{Acknowledgements}
The Institute for Particle Physics Phenomenology at Durham University, UK, its staff and students are gratefully aknowledged for providing an ideal working environment for much of this study.  DJW gratefully acknowledges the hospitality of Jefferson Laboratory in finalising this work. This paper has in part been authored by Jefferson Science Associates, LLC under U.\ S.\ DOE Contract No. DE-AC05-06OR23177. 
This work was also supported by the U.\ S.\ Department of Energy, Office of Nuclear Physics, Contract No.~DE-AC02-06CH11357. We would like to thank Adnan Bashir, Ian Cl\"{o}et, Javier Cobos-Martinez, Craig Roberts, Peter Tandy and Richard Williams for useful discussions.

\appendix

\section{Numerical Method}
\label{app:num}

The equations are solved in Euclidean space using the well known method set out in~\cite{Bloch:1995dd}. Using the standard spherical four-dimensional coordinate system the two inner angles are symmetric for the propagator integrations so may be integrated out. We are then left with the outermost angle which we define as the angle between the incoming propagator momentum $p$ and the loop coordinate $\ell$.

The integrals are logarithmically divergent and are regularised by integrating to some momentum cutoff $\kappa$. This appears in the renormalization procedure. However the results do not depend on it. The functions are subtracted at $\mu^{2}$ and this point is also used to match on to the perturbative result. That is, above this point the one-loop-resummed perturbative result is used as an extrapolation when required by the loop integrations. 

The functions are represented by Chebychev polynomials. These are particularly useful for both their  interpolation properties and for the iterative procedure. They are mapped onto a logarithmic scale typically in the range $p^{2} \in \left[10^{-8},10^{5}\right]$ although the precise numbers are not important. The last Chebychev zero is mapped to the point $\mu^{2}$ since this point is always exactly reproduced by the interpolation. In the IR then the functions have to be represented by some extrapolation. In finding these finite solutions a constant is sufficient for the ghost dressing function and the $c_{g} p^{2}$ term is appropriate for the gluon dressing, where $c_{g}$ is calculated to match smoothly onto the lowest point represented by the Chebychev polynomials.

In performing the iterations we use both the Newton-Raphson and natural iterative procedures. For the finite ghost solutions, then natural iterative procedures are adequate for finding solutions in both equations. A good starting point is always useful and often a requirement, particularly at large couplings. We use Eq.~(\ref{eq_gluonmodel}) for the gluon starting point and $\Gh(p^{2})=1$ is a sufficient starting point for the finite ghost dressing.

\section{Formulae}

\subsection{The Ghost Equation}

The renormalised ghost equation used is,
\begin{align}
&\Gh(p^{2})^{-1}=\tilde{Z}_{3}(\mu^{2},\kappa^{2})\;+g^{2}N_{c}\int \frac{d^{4}\ell}{(2\pi)^{4}}\times\nonumber\\
&\alpha(-\ell_{-},p,\ell_{+})\ \mathcal{K}(p,\ell_{+},\ell_{-})\ \Gh(\ell_{+})\Gl(\ell_{-})
\end{align}
where $\ell_{\pm}=\ell\pm p/2$ where $\ell$ is the loop integration momentum, $\tilde{Z}_{3}$ is the ghost renormalization constant. The function $\alpha$ multiplies the $q_{\mu}$ term in the ghost-gluon vertex, as defined in Eq.~(\ref{eq_gccgen}). The function $\mathcal{K}$ arises from the tensor contractions in the Feynman rules and is given by,
\begin{align}
\mathcal{K}(p,\ell_{+},\ell_{-})&=\frac{-1}{p^2\ell_+^2\ell_-^2}\left(p.\ell_{+}-\frac{p.\ell_-\,\ell_+.\ell_-}{\ell_-^2}\right)\label{eq_ghostK}\\
&=-\frac{\ell^2\sin^2\theta}{\ell_+^2\ell_-^4}\nonumber
\end{align}
where $\theta$ is the integration angle defined via $\ell.p=|\ell||p|\cos\theta$. This equation is typically subtracted from itself at $\mu^{2}$ which is the renormalization point. This removes the term $\tilde{Z}_{3}$.

\subsection{The Gluon Equation}
As described in the text the gluon equation is made up of more terms and contains two dressed one-loop integrations in the formulation we have used here. We give no details of the tadpole term since it yields results that are proportional to the $g_{\mu\nu}$ term of the propagator and hence does not contribute with the method we describe above. The gluon equation that we have used is given by,
\begin{widetext}
\begin{align}
\Gl(p^{2})^{-1}=&{Z}_{3}(\mu^{2},\kappa^{2})+\Pi_{2g}(p^{2},\kappa^{2})+\Pi_{2c}(p^{2},\kappa^{2})\\
				    =&\Gl(\mu^{2})^{-1}+\Pi_{2g}(p^{2},\kappa^{2})-\Pi_{2g}(\mu^{2},\kappa^{2})
				                       +\Pi_{2c}(p^{2},\kappa^{2})-\Pi_{2c}(\mu^{2},\kappa^{2})\\
				    =&\Gl(\mu^{2})^{-1}+\Pi^{\mathrm{sub}}_{2g}(p^{2},\mu^{2})+\Pi^{\mathrm{sub}}_{2c}(p^{2},\mu^{2}).
\end{align}
\end{widetext}
where in the second line we subtract the equation from itself at the point $\mu^{2}$ and in the third line we introduce the subtracted loop integrations $\Pi^{\mathrm{sub}}_{j}(p^{2},\mu^{2})=\Pi_{j}(p^{2},\kappa^{2})-\Pi_{j}(\mu^{2},\kappa^{2})$ which for each diagram are actually all the same in a properly renormalised system when everything above the point $\mu^{2}$ is assumed to be described by one-loop perturbation theory, so we drop the $(\mathrm{sub})$ notation below. We next define the content of the loop integrations in the $\Pi$ functions. We give this only for the most complicated case of vertices $\Gamma^{(2)}$ and $\Gamma^{(C)}$ since the others may be straightforwardly deduced from these by setting factors to $1$ and/or multiplying by the appropriate dressings of the vertex.

First we give the ghost-loop contribution. This contains the two functions $\alpha(k,p,q)$ and $\beta(k,p,q)$ from the ghost-gluon vertex dressings of Eq.~(6) that differ between $\Gamma^{(1)}_{\mu}$ and $\Gamma^{(2)}_{\mu}$,
\begin{widetext}
\begin{align}
\Pi_{2c}(p^2,\mu^2)=
                  &\left(-1\right)\frac{N_c\ g^2(\mu^2)}{(d-1)}\int \frac{d^4\ell}{(2\pi)^4}\frac{\Gh(\ell_+^2,\mu^2)\ \Gh(\ell_-^2,\mu^2)}{p^2\ell_+^2\ell_-^2}\times\nonumber\\ 
                        &\Bigl(\Bigr.\alpha(-p;\ell_-,\ell_+)\,\mathcal{M}_{\alpha}(p,\ell)\,+\,\beta(-p;\ell_-,\ell_+)\,\mathcal{M}_{\beta}(p,\ell)\Bigl.\Bigr),
\label{eq_gluonsde_set1}
\end{align}
where $d=4$ is the number of dimensions, $g$ is the value of the coupling at the renormalization point, the functions $\alpha$ and $\beta$ arise from the two terms in the ghost-gluon vertex, Eq.~(6). The kinematic terms are contracted together into the two $\mathcal{M}$ functions, which are derived to be, 
\begin{align}
\mathcal{M}_{\alpha}(p,\ell)&= \frac{-1}{p^{2}}\left(p^2\left[\ell^2-\frac{p^2}{4}\right]-d\left[(\ell.p)^2-\frac{p^4}{4}\right]\right)\label{eq_Mbare},\\
   \mathcal{M}_\beta(p,\ell)&= (1-d)\,p.\ell_-.
\end{align} 
For the bare vertex, the $\beta$ term is zero and for the transverse vertex, $\Gamma^{(1)}$, then $\alpha(k,p,q)=1$. For the Taylor-respecting vertex $\Gamma^{(2)}$ then we replace $p_{\mu}\to k_{\mu}+q_{\mu}$ in Eq.~(\ref{eq_gcctt}) and read off the respective coefficients. Similarly for the gluon loop we have, 
\begin{align}
\Pi_{2g}(p^2)=&\left(\frac{1}{2}\right)\,\frac{N_c\ g^2(\mu^2)}{(d-1)}\int \frac{d^4\ell}{(2\pi)^4}\frac{\Gl(\ell_+^2)\, \Gl(\ell_-^2)}{4\,(p^2\ell_+^2\ell_-^2)^2}\Bigl(\mathcal{R}_{1}\mathcal{Q}_{1}+\mathcal{R}_{2}\mathcal{Q}_{2}+\mathcal{R}_{3}\mathcal{Q}_{3}\Bigr)
\end{align}
where the ratios of dressing functions are chosen to be,
\begin{align}
\mathcal{R}_{1}=\frac{1}{2}\left(\frac{\Gh(\ell_{-}^{2})}{\Gl(\ell_{+}^{2})}+\frac{\Gh(\ell_{-}^{2})}{\Gl(p^{2})}\right)\quad\quad
\mathcal{R}_{2}=\frac{1}{2}\left(\frac{\Gh(p^{2})}{\Gl(\ell_{+}^{2})}+\frac{\Gh(p^{2})}{\Gl(\ell_{-}^{2})}\right)\quad\quad
\mathcal{R}_{3}=\frac{1}{2}\left(\frac{\Gh(\ell_{+}^{2})}{\Gl(\ell_{-}^{2})}+\frac{\Gh(\ell_{+}^{2})}{\Gl(p^{2})}\right)
\end{align}
and the associated momentum contractions are,
%
%
\begin{align}
\mathcal{Q}_{1}&=2(\ell.p^2-\ell^2p^2)\left[4\ell^2(5p^2+2d\ell.p)+p^2(7p^2+2(12-d)\ell.p)\right]\,,\\
\mathcal{Q}_{2}&=-32d(\ell.p)^4 + (3d-4)p^4(\ell.p)^2-48\ell^6 p^2+24\ell^4(2d\ell.p^2-p^4)+\ell^2 p^2(8(d+6)\ell.p^2+p^4)\,,\\
\mathcal{Q}_{3}&=2(\ell.p^2-\ell^2p^2)\left[4\ell^2(5p^2-2d\ell.p)+p^2(7p^2-2(12-d)\ell.p)\right]\,.
\end{align}
\end{widetext}
In order to obtain a vertex proportional to the bare Lorentz structure as in $\Gamma^{(A)}$ and $\Gamma^{(B)}$ then we just have to make a replacement, for example for $\Gamma^{(A)}$ we have,
\begin{align}
\mathcal{R}_{i}\to\frac{\Gh(\ell_{+}^{2})\Gh(\ell_{-}^{2})}{\Gl(\ell_{+}^{2})\Gl(\ell_{-}^{2})},
\end{align}
for all three $\mathcal{R}_{i}$'s, and a similar substitution is made to obtain the equation for the $\Gamma^{(B)}$ dressing.

The smoothed step function used for interpolating between IR and UV vertex functions is taken from~\cite{Fischer:2008uz}. However, a range of forms may be used,
\begin{equation}
\mathcal{F}_{\mathrm{IR}}(k,p,q)=\frac{\Lambda_{\mathrm{IR}}^{6}}{(k^{2}+\Lambda_{\mathrm{IR}}^{2})(p^{2}+\Lambda_{\mathrm{IR}}^{2})(q^{2}+\Lambda_{\mathrm{IR}}^{2})}
\label{eq_fir}
\end{equation}
where $\Lambda_{\mathrm{IR}}$ gives the change-over between the perturbative and non-perturbative form. The solutions are sensitive to the choice of this parameter. In the modified form of the vertex of Eq.~(\ref{eq_gcctt}) an extra parameter $\mathcal{N}_{\mathrm{IR}}$ multiplying the non-perturbative term was introduced.

\subsection{Resummed One-loop running}

In the UV we match the gluon and ghost propagator dressings on to the resummed one-loop form from perturbation theory. This is necessary as some extrapolation beyond the Chebychev region in the UV is always required. The form that we find to be most useful numerically is given by,
\begin{align}
\Gl(p^2)&=\Gl(\mu^2)\left(\alpha(\mu^2)/\alpha(p^2)\right)^{d_g}\\
\Gh(p^2)&=\Gh(\mu^2)\left(\alpha(\mu^2)/\alpha(p^2)\right)^{d_c}\; ,
\end{align}
where $d_g\,=\,-13/22$ ({\it cf.} the modeling in Eq.~(13)) and $d_c\,=\,-9/44$ (as appropriate for the pure gauge sector)
which is applied where $p^{2}>\mu^{2}$. Numerically, we also use the standard relation for the running coupling,
\begin{align}
\frac{1}{\alpha(p^{2})}&=\frac{1}{\alpha(\mu^{2})}+\frac{11N_{c}}{12\pi}\log\frac{p^{2}}{\mu^{2}}.
\end{align}
where $\alpha(\mu^{2})$ is specified to fix the scale.

Since only the triple-gluon vertex $\Gamma^{(A)}_{\mu\nu\rho}$ perfectly reproduces the perturbatively resummed one-loop running, it is useful to interpolate between this in UV and the other vertices in the mid-momentum region and below. The WSTI triple-gluon vertex, $\Gamma^{(C)}_{\mu\nu\rho}$ in particular induces a steeper gluon dressing function than would be expected from perturbative studies. Some deviation is allowed due to the effects of higher orders perturbatively and since we are still at the modeling stage then interpolating between these two vertices between an IR scale $\Lambda_{\mathrm{IR}}^{2}$ and $\mu^{2}\simeq M_{Z}^{2}$ is currently the best that can be done. The transition is implemented here using $\mathcal{F}_{\mathrm{IR}}$ (see Eq.~(\ref{eq_gcctransv}) and Eq.~(\ref{eq_fir})) and $(1-\mathcal{F}_{\mathrm{IR}})$ to multiply the respective terms.

\bibliographystyle{apsrev4-1}
\bibliography{sderefs2}

\end{document}